\documentclass[english, preprint, amsmath,amssymb, aip]{revtex4-2}
\usepackage[T1]{fontenc}
\usepackage{float}
\usepackage{amsmath}
\usepackage{amssymb}
\usepackage{enumitem}
\usepackage{xcolor}
\usepackage{graphicx}
\usepackage{dcolumn}
\usepackage{bm}
\usepackage{hyperref}
\usepackage{setspace}
\usepackage{threeparttablex}
\usepackage{booktabs}
\usepackage{array}
\usepackage[title]{appendix}

\begin{document}

\title{Unraveling Abnormal Collective Effects via the Non-Monotonic Number Dependence of Electron Transfer in Confined Electromagnetic Fields}

\author{Shravan Kumar Sharma}
\affiliation{Department of Chemistry and Biochemistry, 251 Nieuwland Science Hall, Notre Dame, Indiana 46556, United States}
\author{Hsing-Ta Chen}
\email{hchen25@nd.edu}
\affiliation{Department of Chemistry and Biochemistry, 251 Nieuwland Science Hall, Notre Dame, Indiana 46556, United States}

\date{\today}

\begin{abstract}
Strong light-matter coupling within an optical cavity leverages the collective interactions of molecules and confined electromagnetic fields, giving rise to the possibilities of modifying chemical reactivity and molecular properties. While collective optical responses, such as enhanced Rabi splitting, are often observed, the overall effect of the cavity on molecular systems remains ambiguous for a large number of molecules.
In this paper, we investigate the non-adiabatic electron transfer (ET) process in electron donor-acceptor pairs influenced by collective excitation and local molecular dynamics. Using the timescale difference between reorganization and thermal fluctuations, we derive analytical formulas for the electron transfer rate constant and the polariton relaxation rate. These formulas apply to any number of molecules ($N$) and account for the collective effect as induced by cavity photon coupling. 
Our findings reveal a non-monotonic dependence of the rate constant on $N$, which can be understood by the interplay between electron transfer and polariton relaxation. As a result, the cavity-induced quantum yield increases linearly with $N$ for small $N$ (as predicted by a simple Dicke model), but shows a turnover and suppression for large $N$ (consistent with the large $N$ problem of polariton chemistry). 
We also interrelate the thermal bath frequency and the number of molecules, suggesting the optimal number for maximizing enhancement. 
The analysis provides an analytical insight for understanding the collective excitation of light and electron transfer, helping to predict the optimal condition for effective cavity-controlled chemical reactivity.
\end{abstract}
\maketitle

\newpage
\section{Introduction}

Polaritons, hybrid light-matter states emerging from strong coupling between matter excitations and confined photons, have drawn a lot of attention for their promising potential for a broad spectrum of applications, ranging from modifying molecular reactions to controlling transport and optical properties of materials.\cite{ebbesen_hybrid_2016,garcia-vidal_manipulating_2021,xiong_molecular_2023,xiang_molecular_2024,xu_ultrafast_2023,lee_controlling_2024,balasubrahmaniyam_enhanced_2023,weight_cavity_2024,li_molecular_2022}
Significant modulation of chemical reactivity has been observed when the cavity mode is tuned to be resonant with certain vibration modes of reactant molecules in the gas phase or a solution.\cite{ahn_modification_2023,xiang_molecular_2024,wright_versatile_2023,chen_exploring_2024}
Enhanced transport mobility of charge carriers has been reported showing long-range interactions of polariton as induced by coherent mixing with light.\cite{balasubrahmaniyam_enhanced_2023,lee_controlling_2024,xu_ultrafast_2023,suyabatmaz_vibrational_2023,rozenman_long-range_2018}
To understand these intriguing phenomena, many new theoretical methods are developed to account for the coherent mixing of light and matter using quantum electrodynamics (QED). \cite{sidler_perspective_2022,mandal_theoretical_2023,mandal_investigating_2019}
That being said, most of current methods often focus on a single-molecule behavior and extrapolate the collective interactions of multiple molecules.

To account for the collective effect, an idealized model considers a molecular ensemble comprising $N$ identical molecules, and their vibrational or electronic excitations are coherently coupled to a shared cavity mode with the same coupling strength $g_c$, termed the Tavis-Cummings model.\cite{tavis_exact_1968}
This idealized model predicts that the collective interactions lead to an enhanced light-matter coupling strength by a factor of $\sqrt{N}$.
Such enhancement often manifests in the optical response of the molecular ensemble, increasing the Rabi splitting (the energy difference between polaritonic states) by a factor of $N$.
This monotonic relation between the effective coupling strength and the Rabi splitting enhancement has been widely used to determine the number of molecules that are involved in the polaritonic states.\cite{shalabney_coherent_2015,wright_versatile_2023,xiong_molecular_2023,ahn_modification_2023,schwennicke_extracting_2024}
Regarding the chemical reaction rate modification, incorporating the collective interactions presents great challenges.
A straightforward approach is to use the effective coupling strength (as obtained by the collective optical response) for simulating the local reaction dynamics in a single molecule model system.
Very often, this approach overestimates the collective effect on the modified reaction rate and does not account for the inherent disorder of molecular ensembles.\cite{campos-gonzalez-angulo_polaritonic_2020,zhou_interplay_2023} 
More sophisticated numerical approaches are developed for \emph{ab initio} QED simulations, but usually limited to a small number of molecules.\cite{ruggenthaler_quantum-electrodynamical_2018,schafer_modification_2019,jestadt_light-matter_2019,schafer_relevance_2020,sidler_perspective_2022} 

In the large $N$ regime, an important issue remains unclear: how the collective light-matter interactions (which involve many molecules) modify chemical reactions on a single molecule?
To see this, we consider the scenario where the observed reaction occurs on a single molecule and the rest of $N-1$ molecules remain non-reactive. 
In this case, while the polariton states energy difference is enhanced by $\sqrt{N}$, the effective coupling strength is shared among all $N$ molecules and the reacting molecule only experiences $1/N$ part of the effective coupling as induced by the cavity.
Consequently, the advantages offered by the collective interactions are often offset by the presence of an overwhelmingly large number ($N-1$) of non-reactive molecules, which are considered in the optically dark state.\cite{martinez-martinez_triplet_2019,du_vibropolaritonic_2023,hu_quasi-diabatic_2022}
This issue is referred to as the polariton ``large $N$ problem,'' implying that the collective effect on the local reaction cannot increase monotonically with $N$, especially in the large $N$ regime. 
Huo et al formulated the vibrational strong coupling (VSC) modified reaction rate constant under the resonant condition using Fermi's golden rule.\cite{ying_resonance_2024} 
Yuen-Zhou et al developed the collective dynamics using truncated equations (CUT-E) to systematically address this issue in the context of the energy transfer mechanism.\cite{perez-sanchez_simulating_2023} 
Despite many efforts to understand the collectivity of strong coupling phenomena, how forming molecular polariton impacts electron transfer reactions is less studied and still not fully understood.

Here we remark that strong light-matter coupling between molecules and confined electromagnetic fields can occur in different scenarios.  
Vibrational strong coupling describes a molecular vibrational mode that is near-resonant with a cavity mode. The modification of the ground-state reaction rates can be captured by propagating the classical motion of nuclei and cavity photon mode as coupled harmonic oscillators on an electronic ground surface.\cite{li_cavity_2020,li_polariton_2022}
Electronic strong coupling (ESC) refers to a local electronic excitation with a large radiative transition dipole moment that is near-resonant with a cavity mode. For example, polariton-mediate electron transfer (PMET) supposes that the confined electric field is strongly coupled to the local excitation between the ground state and donor state.\cite{mandal_polariton-mediated_2020}
Electron transfer strong coupling (ETSC) presumes that the transition energy of a charge transfer state is near-resonant with a cavity photon excitation. In this case, photon absorption or emission modulates the electronic transition between the donor and acceptor states.\cite{campos-gonzalez-angulo_resonant_2019,semenov_electron_2019}

Focusing on the collective ETSC reaction rate modification, we notice that, in addition to the photon-modulated electronic transition, the non-adiabatic electron transfer rate is governed by the following local mechanisms:
(a) the reorganization process of the nuclear degrees of freedom, which is the driving force of the non-adiabatic ET in the absence of cavity photon,\cite{nitzan_chemical_2006}
(b) thermal fluctuations stemming from homogeneous disorder of the environment, which dissipate the electronic coherence and relax the local nuclear motions,\cite{cui_collective_2022,climent_kubo-anderson_2024}
(c) the dipole self-energy as induced by the overall back-action of the molecular polarization on the photon field, which explicitly hybridizes the electronic and photonic degrees of freedom.\cite{mandal_polariton-mediated_2020}
The interplay between these local effects and the photon-modulated electronic transitions for a single donor-acceptor pair has been investigated by Nitzan et al, showing a significant enhancement of the reaction rate constant when the potential energy surfaces become barrier-less.\cite{semenov_electron_2019}  
However, when considering multiple molecules, the simplest generalization of this model predicts the enhancement factor to be proportional to $N$.

In this paper, we employ the Holstein-Tavis-Cummings (HTC) model\cite{zhang_collective_2021,ying_theory_2024} to capture the non-monotonic $N$-dependence of the ETSC reaction rate using the Marcus theory framework.
We aim to address the collective ETSC effect on both the electron transfer rate constant and the polariton relaxation as induced by the nuclear bath degrees of freedom.
This paper is organized as follows.
In Sec.~\ref{sec:model}, we formulate the HTC model Hamiltonian for a molecular ensemble confined within an optical cavity and treat local nuclear motions and thermal fluctuations using semiclassical approximations. 
In Sec.~\ref{sec:method}, we separate the local nuclear modes into the reorganization and thermal fluctuation modes and derive the analytical expressions for the non-adiabatic ET rate and polariton relaxation rates. 
In Sec.~\ref{sec:result}, we present the results focusing on $N$-dependence of the ET rate constant and polariton relaxation, as well as the quantum yield induced by collective ETSC.
Finally, we discuss our findings and conclude in Sec.~\ref{sec:discussion}.

Throughout the paper, we denote a hat operator (such as $\hat{H}$) as a quantum operator and a bold-faced notation (such as $\mathbf{H}$) as the matrix representation. We use $\vec{r}$ for a three-dimensional vector in real space and a curly bracket $\{R\}=\{R_j|j=1,\cdots,N\}$ for a set of variables.
 
\section{Model Hamiltonian}\label{sec:model}
We consider an ensemble of $N$ molecules interacting with the electromagnetic field confined within an optical cavity. 
The total Hamiltonian takes the form 
\begin{equation}
    \hat{H}=\hat{H}_{\text{C}}+\hat{H}_{\text{I}}+\sum_{j=1}^N\hat{H}_{\text{M},j}
\end{equation}
where $\hat{H}_{\text{C}}$ is the cavity photon Hamiltonian, $\hat{H}_{\text{M},j}$ represents the $j$-th molecular subsystem, and $\hat{H}_{\text{I}}$ describes the light-matter interactions.
For the confined electromagnetic field in the optical cavity, we assume the cavity supports a quantized photon mode of frequency $\omega_c$ and disregards other photon modes and radiation loss of the cavity.
Explicitly, the cavity photon Hamiltonian is given by 
$\hat{H}_{\text{C}}=\hbar\omega_c\hat{a}^\dagger\hat{a}$ where $\hat{a}^\dagger$($\hat{a}$) is the creation (annihilation) operator of the cavity photon mode and we neglect the zero-point energy.

\subsection{Molecular Hamiltonian}
The molecular subsystem is comprised of a set of donor-acceptor pairs that do not interact with each other directly. 
For each pair, we describe the electron transfer (ET) process using the spin-boson-type model \cite{nitzan_chemical_2006}
\begin{equation}
\begin{split}
    \hat{H}_{\text{M},j}=&
    E_{D}|D_j\rangle\langle D_j|\\
    &+\left(E_{A}+\sum_{\alpha}c_{\alpha,j}(\hat{b}_{j,\alpha}^{\dagger}+\hat{b}_{j,\alpha})\right)|A_j\rangle\langle A_j| \\ 
    &+H_{AD}|A_j\rangle\langle D_j|+H_{DA}|D_j\rangle\langle A_j|\\
    &+\sum_{\alpha}\hbar\omega_{\alpha}(\hat{b}_{j,\alpha}^{\dagger}\hat{b}_{j,\alpha}+\frac{1}{2})
\end{split}
\end{equation}
Here $|D_j\rangle$ and $|A_j\rangle$ are the diabatic electronic states of the $j$-th donor--acceptor pair with the excess electron located on the donor and the acceptor, respectively, and $E_{D}$ and $E_{A}$ are the energy of the corresponding electronic states.
The electronic coupling between the diabatic states is $H_{AD}=\langle A_j|\hat{H}_{\text{M},j}|D_j\rangle=H_{DA}^*$.
For each donor--acceptor pair, the electronic states are coupled to a boson bath, modeling the nuclear motions and the solvent environment. 
Note that, since we assume the donor-acceptor pairs are identical, $E_{D}$, $E_{A}$, and $H_{AD}$ do not depend on $j$.

For the nuclear degrees of freedom, we denote $\hat{b}_{j,\alpha}^{\dagger}(\hat{b}_{j,\alpha})$ as the creation (annihilation) operator of the boson mode of frequency $\omega_{\alpha}$ associated with the $j$-th molecule.
Without loss of generality, we assume that the boson bath has a linear coupling to only the acceptor electronic state with the coupling strength $c_{\alpha,j}$, characterizing the interaction between the acceptor state and the boson modes. 
Furthermore, due to the timescale difference between the electronic transitions and nuclear motions, we treat the nuclear degrees of freedom as a set of classical harmonic oscillators by approximating $\frac{1}{\sqrt{2\hbar\omega_{\alpha}}}(\hat{b}_{j,\alpha}^{\dagger}+\hat{b}_{j,\alpha})\rightarrow q_{j,\alpha}$ and $i\sqrt{\frac{\hbar\omega_{\alpha}}{2}}(\hat{b}_{j,\alpha}^{\dagger}-\hat{b}_{j,\alpha})\rightarrow p_{j,\alpha}$. 
Here $(q_{j,\alpha}, p_{j,\alpha})$ are the position and momentum coordinates of the classical harmonic oscillation mode at frequency $\omega_\alpha$.
Effectively, the classical harmonic oscillators can recast in the form of coupled harmonic oscillators where a primary mode $(R_j,P_j)$ is coupled to the electronic state and also in contact with a thermal bath $\{\tilde{q}_{j,\alpha},\tilde{p}_{j,\alpha}\}$ that does not couple to the electronic states directly. 
Within this semiclassical framework, the molecular Hamiltonian can be expressed in terms of $\hat{H}_{\text{M},j}=\frac{P_j^{2}}{2}+\hat{H}_{j}(R_j)+H_{\text{B},j}(R_j,\{\tilde{q}_{j,\alpha},\tilde{p}_{j,\alpha}\})$ where $H_{\text{B},j}$ represents the thermal bath.
The molecular Hamiltonian $\hat{H}_{j}(R_j)$ is an electronic operator as a function of the primary coordinate $R_j$
\begin{equation}
\begin{split}
\hat{H}_{j}(R_j)=&V_{D}(R_j)|D_j\rangle\langle D_j|+V_{A}(R_j)|A_j\rangle\langle A_j|\\
&+{H}_{AD}|A_j\rangle\langle D_j|+{H}_{DA}|D_j\rangle\langle A_j|
\end{split}
\end{equation}
The potential energy surfaces (PES) of the electronic donor and acceptor states are given by 
\begin{subequations}\label{V_D_and_V_A}
    \begin{align}
    V_{D}(R_j)&=E_{D}+\frac{1}{2}\omega_{v}^{2}R_j^{2}\\
    V_{A}(R_j)&=E_{A}+\lambda_{v}R_j+\frac{1}{2}\omega_{v}^{2}R_j^{2}
    \end{align}
\end{subequations}
where $V_A(R_j)=V_D(R_j)+E_{AD}+\lambda_vR_j$ and $E_{AD}=E_{A}-E_{D}$.
Here  $\lambda_{v}$ is the primary mode coupling strength and $\omega_v$ is the characteristic frequency of the primary mode.
Note that, since we assume the donor-acceptor pairs are identical, $\lambda_{v}$ and $\omega_v$ do not depend on $j$.

The overall effect of the thermal bath can be considered in terms of Langevin dynamics of coupled harmonic oscillators.\cite{nitzan_chemical_2006}
The primary mode coordinate follows Langevin equation, $\ddot{R}_j=-\omega_{v}^{2}R_j-\gamma_v P_j+\eta_j(t)$. 
Here $\gamma_v$ is the damping coefficient and $\eta_j(t)$ is a stochastic random force.
We assume the stochastic force $\eta_j(t)$ to be a Markovian Gaussian random variable following $\langle\eta_j(t)\rangle_T=0$ and $\langle\eta_j(t)\eta_j(0)\rangle_T=2\gamma_v k_BT\delta(t)$ where $\langle\cdots\rangle_T$ denote the thermal average of the nuclear degrees of freedom. 
Note that, while $\eta_j(t)$ represents local fluctuation of the $j$-th molecule, other parameters ($\gamma_v$ and $T$) do not depend on $j$ as we consider homogeneous disorder.

In summary, within the semiclassical framework, the spin-boson model of each donor-acceptor pair is characterized by the following parameters: 
(\textit{1}) $E_{D}$ and $E_{A}$ are the diabatic energies of the donor and acceptor states, 
(\textit{2}) $H_{AD}$ is the electronic coupling between the donor and acceptor,
(\textit{3}) $\lambda_{v}$ is the system-bath coupling strength,
(\textit{4}) $\omega_v$ is the characteristic frequency of the bath, 
(\textit{5}) $\gamma_v$ is the damping coefficient of the primary mode,
(\textit{6}) $T$ is the temperature of the thermal bath.
For simplicity, we suppose that all the donor-acceptor pairs of the molecular ensemble are identical (i.e. these parameters do not depend on $j$) and disregard the inter-molecular interactions. 
In the end, the molecular Hamiltonian of each donor-acceptor pair is expressed in the matrix form (in the basis of the diabatic states $|D_j\rangle$ and $|A_j\rangle$) 
\begin{equation}
    \hat{H}_{\text{M},j}=\frac{P_j^{2}}{2}+
    \left[\begin{array}{cc}
    V_{D}(R_j) & H_{DA} \\
    H_{AD} & V_{A}(R_j) 
    \end{array}\right]
\end{equation}
where the classical nuclear coordinates $R_j$ follow Langevin equation. 

\subsection{Light-matter interactions}
Following Semenov and Nitzan in Ref.~\citenum{semenov_electron_2019}, we describe light-matter interactions between the donor--acceptor pair and the cavity photon mode by the minimal-coupling QED Hamiltonian under the Power--Zienau--Woolley (PZW) gauge transformation and the long-wavelength approximation.\cite{cohen-tannoudji_photons_1997} 
The light-matter interaction Hamiltonian includes the dipole-electric field coupling ($\hat{H}_{\text{DE}}$) and the dipole self-energy term ($\hat{H}_{\text{DS}}$) 
\begin{equation}
\hat{H}_{\text{I}}=\hat{H}_{\text{DE}}+\hat{H}_{\text{DS}}=\sum_j\vec{\mu}_j\cdot\vec{E}(\vec{r}_j)+\sum_j\frac{\left(\vec{\mu}_j\cdot\vec{\xi}\right)^2}{2\epsilon_0{\cal V}}
\end{equation}
where $\epsilon_0$ is the vacuum permittivity, $\cal V$ is the cavity volume. 
Here $\vec{\mu}_j=-e\vec{r}_j$ is the dipole operator and $\vec{r}_j$ is the position operator of the excess electron in the $j$-th donor--acceptor pair (where the molecular subsystem excluding the excess electron is assumed to be charge-neutral).
The electromagnetic field of the cavity photon mode at the position $\vec{r}_j$ is given by $\vec{E}(\vec{r}_j)=i\vec{\xi}\sqrt{\frac{\omega_c}{2{\cal V}\epsilon_0}}(\hat{a}e^{i\vec{k}\cdot\vec{r}_j}-\hat{a}^{\dagger}e^{-i\vec{k}\cdot\vec{r}_j})$ where $\vec{\xi}$ is the photon mode polarization vector associated with the wave vector $\vec{k}$ and $\omega_c=c|\vec{k}|$.

In terms of the electronic diabatic states ($|D_j\rangle$ and $|A_j\rangle$), the electric dipole coupling Hamiltonian takes the form
\begin{equation}\label{H_dip}
\begin{split}
    \hat{H}_{\text{DE}}=&\hbar\omega_{c}(\hat{a}-\hat{a}^{\dagger})\times\\
    &\sum_j\bigg[g_{D}|D_j\rangle\langle D_j|+g_{A}|A_j\rangle\langle A_j|\\
    &+t_{AD}|A_j\rangle\langle D_j|+t_{DA}|D_j\rangle\langle A_j|\bigg]
\end{split}
\end{equation}
Here the coupling strengths are defined by the following dimensionless parameters 
\begin{align}
    g_{D}&=i\sqrt{\frac{1}{2\hbar\omega_{c}\Omega\epsilon_{0}}}\vec{\xi}\cdot\vec{d}_{DD}\\
    g_{A}&=i\sqrt{\frac{1}{2\hbar\omega_{c}\Omega\epsilon_{0}}}\vec{\xi}\cdot\vec{d}_{AA}\\
    t_{AD}&=i\sqrt{\frac{1}{2\hbar\omega_{c}\Omega\epsilon_{0}}}\vec{\xi}\cdot\vec{d}_{AD}
\end{align}
and $t_{DA}=-t_{AD}^*$.
Here we consider the idealized scenario of $N$ identical molecules equally coupled to the cavity photon mode.
This idealized scenario involves the following assumptions. 
First, we neglect the spatial and orientational disorder of the molecular dipole moments (i.e. all the molecules have identical dipole moments, $\vec{d}_{DD}=\langle D_j|\vec{\mu}_j|D_j\rangle$, $\vec{d}_{AA}=\langle A_j|\vec{\mu}_j|A_j\rangle$, and $\vec{d}_{AD}=\langle A_j|\vec{\mu}_j|D_j\rangle$, $\vec{d}_{DA}=\vec{d}_{AD}^*$ for all $j=1,\cdots,N$). 
Second, we employ the long-wavelength approximation --- assuming that the wavelength of the cavity photon mode is much larger than the spatial extent of the molecular ensemble, all the molecules experience the same electric field $\vec{E}=i\vec{\xi}\sqrt{\frac{\omega_c}{2\Omega\epsilon_0}}(\hat{a}-\hat{a}^{\dagger})$.

Following these assumptions, the dipole self-energy term can be expressed in terms of the electronic diabatic states ($|D_j\rangle$ and $|A_j\rangle$) 
\begin{equation}\label{H_dse}
\begin{split} \hat{H}_{\text{DS}}=&\hbar\omega_{c}\sum_j\bigg[|g_{D}|^{2}|D_j\rangle\langle D_j|+|g_{A}|^{2}|A_j\rangle\langle A_j|+|t_{DA}|^{2}\bigg]
    \\
    &-\hbar\omega_{c}(g_{D}+g_{A})\sum_j\bigg[t_{AD}|A_j\rangle\langle D_j|+t_{DA}|D_j\rangle\langle A_j|\bigg].
\end{split}
\end{equation}
which consists of the quadratic order of the coupling strength parameters.
Note that Eq.~\eqref{H_dse} indicates that the dipole self-energy term does not involve photon absorption and emission, implying that we can consider $\hat{H}_{\text{DS}}$ as a local molecular Hamiltonian. 
Since $\hat{H}_{\text{DS}}$ contains the quadratic order of the coupling strength parameters, its impacts become significant in the strong coupling regime.\cite{schafer_relevance_2020,taylor_resolution_2020}
We emphasize that the second term in $\hat{H}_{\text{DS}}$ has an effective contribution to the diabatic coupling between the donor and acceptor states, which also impacts the electron transfer process.

\subsection{Polaron transformation}
Next, we employ the polaron-type unitary transformation to simplify the total Hamiltonian. 
Following Ref.~\citenum{semenov_electron_2019}, the polaron transformation of the Hamiltonian takes the form $\tilde{H}=e^{\hat{\cal S}}\hat{H}e^{-\hat{\cal S}}$ where $\hat{\cal S}$ is chosen to be 
\begin{equation}\label{S-operator}
    \hat{\cal S}=\sum_j \left(g_{D}^{*}\hat{a}^{\dagger}-g_{D}\hat{a}\right)|D_j\rangle\langle D_j|+\left(g_{A}^{*}\hat{a}^{\dagger}-g_{A}\hat{a}\right)|A_j\rangle\langle A_j|
\end{equation}
Under the polaron transformation, the cavity Hamiltonian remains unchanged $\tilde{H}_\text{C}={H}_\text{C}$ and the molecular Hamiltonian becomes,\footnote{The unitary transformation defined by Eq.~\eqref{S-operator} is a generalization of Eq.~(18) in Ref.~\citenum{semenov_electron_2019} for $N$ non-interacting molecules. The evaluation of the matrix elements and the Hamiltonian is straightforward based on Appendix B and C in Ref.~\citenum{semenov_electron_2019}. Therefore we do not repeat the same process in this paper.} 
\begin{equation}\label{Hmol_polaron}
    \tilde{H}_{\text{M},j}=\frac{P_j^{2}}{2}+
    \left[\begin{array}{cc}
    V_{D}(R_j) & H_{DA}\hat{\cal F} \\
    H_{AD}\hat{\cal F} & V_{A}(R_j) 
    \end{array}\right]
\end{equation}
Here the PES does not change and the diabatic coupling term is dressed by the photonic shift operator
\begin{equation}
    \hat{\cal F}=\exp\left\{ g_{AD}^{*}\hat{a}^{\dagger}-g_{AD}\hat{a}\right\}
\end{equation}
where $g_{AD}=g_{A}-g_{D}$ is proportional to the position difference of the excess electron located at the donor and acceptor states, $g_{AD} \propto (\vec{d}_{AA}-\vec{d}_{DD})=-e (\vec{r}_{A}-\vec{r}_{D})$. 
The light-matter interaction Hamiltonian (Eq.~\eqref{H_dip} and Eq.~\eqref{H_dse}) is transformed into the following form 
\begin{equation}\label{HdE_polaron}
\begin{split}
    \tilde{H}_{\text{I}}=\hbar\omega_{c}
    \sum_j&\hat{\cal F}\left[t_{AD}g_{AD}+t_{AD}(\hat{a}-\hat{a}^{\dagger})\right]|A_j\rangle\langle D_j|\\
    +& \left[-t_{DA}g_{DA}+t_{DA}(\hat{a}-\hat{a}^{\dagger})\right]\hat{\cal F}^{\dagger}|D_j\rangle\langle A_j|
\end{split}
\end{equation}
Note that, by design, the polaron transformation eliminates the quadratic coupling terms of the light-matter interaction Hamiltonian.
In the end, the polaron-transformed Hamiltonian can be expressed in the form of the standard coupled electron-nuclei Hamiltonian $\tilde{H}=\sum_j\frac{P_j^{2}}{2}+\tilde{H}_\text{eff}(\{R\})$ where 
\begin{equation}\label{Hall_polaron}
\begin{split}
    \tilde{H}_\text{eff}(\{R\})=&\hat{H}_{\text{C}}+\sum_j
    \left[\begin{array}{cc}
    V_{D}(R_j) & \hat{\cal H}_{DA} \\
    \hat{\cal H}_{AD} & V_{A}(R_j) 
    \end{array}\right]
\end{split}
\end{equation}
and the effective diabatic coupling is 
\begin{align}
    \hat{\cal H}_{AD}&\equiv \hat{\cal F}\left[H_{AD}+\omega_{c}t_{AD}g_{AD}+\omega_{c}t_{AD}(\hat{a}-\hat{a}^{\dagger})\right]
\end{align}
and $\hat{\cal H}_{DA}=\hat{\cal H}_{AD}^\dagger$.
We emphasize that the effective coupling becomes a photonic operator and the diabatic state energy surfaces $V_{D}(R_j)$ and $V_{A}(R_j)$ are unchanged with the polaron transformationthe.

\subsection{The dressed state representation}
Now we expand the polaron-transformed Hamiltonian in terms of the dressed states denoted as $|\psi_1,\cdots,\psi_N\rangle\otimes|m\rangle $.
Here $|\psi_1,\cdots,\psi_N\rangle$ is the combination of the electronic states for $\psi_j=D_j, A_j$ indicating the electronic state of the $j$-th molecule, and $|m\rangle$ is the photon number state following $\hat{a}^{\dagger}\hat{a}|m\rangle=m|m\rangle$ where $m=0,1,\cdots$ is the number of photon within the cavity.
For convenience, we denote $|\{D\}\rangle\equiv|D_{1}\cdots D_{N}\rangle$ as the total electronic state where the excess electrons of all the molecules are on the donor states, and $|\{A_{j}\}\rangle\equiv|D_{1}\ldots A_{j}\ldots D_{N}\rangle$ is the electronic state where the excess electron of $j$-th molecule is in the acceptor state while all other molecules stay in the donor state. 
Particularly, we focus on the single electron transfer under the influence of collective light-matter interactions and denote the dressed state basis as $|\{D\}m\rangle=|\{D\}\rangle\otimes|m\rangle$ and $|\{A_{j}\}m\rangle=|\{A_j\}\rangle\otimes|m\rangle$ for $j=1,\cdots,N$ and $m=0,1,2,\cdots$.

Expanding Eq.~\eqref{Hall_polaron} in the dressed states ($|\{D\}m\rangle$ and $|\{A_{j}\}m\rangle$), the diagonal terms are the potential energy surface of $\{D\}$ and $\{A_{j}\}$ shifted by the photon number $m$:
\begin{align}
    V_{\{D\}m}(\{R\})&=\sum_{j=1}^{N}V_{D}(R_{j})+m\hbar\omega_c \\
    V_{\{A_{j}\}m}(\{R\})&=V_{A}(R_{j})+\sum_{j'\neq j}^{N}V_{D}(R_{j'})+m\hbar\omega_c
\end{align}
The effective diabatic coupling terms can be evaluated by 
\begin{equation}
    \begin{split}
    {\cal T}_{lm}
    =&\langle l|\hat{{\cal H}}_{AD}|m\rangle\\
    =&\left(H_{AD}+\hbar\omega_{c}t_{AD}g_{AD}\right)F_{l,m}\\
    &+\hbar\omega_{c}t_{AD}\left(\sqrt{m}F_{l,m-1}-\sqrt{m+1}F_{l,m+1}\right)
    \end{split}
\end{equation}
Here $F_{l,m}=\langle l|\hat{\cal F}|m\rangle$ is the matrix element of the shift operator, which can be evaluated using 
\begin{equation}
    F_{l,m}=\sqrt{\frac{l!}{m!}}(- g_{AD})^{m-l}\exp\left(-\frac{1}{2}|g_{AD}|^{2}\right)L_{l}^{m-l}\left(|g_{AD}|^{2}\right)
\end{equation}
where $L_{l}^{m-l}(x)$ is associated Laguerre polynomial for $l,m\ge0$. 
Explicitly, the first few associated Laguerre polynomials are given by $L_{0}^{k}(x)=1$, $L_{1}^{k}(x)=-x+k+1$, and $L_{2}^{k}(x)=\frac{1}{2}\left[x^{2}-2(k+2)x+(k+1)(k+2)\right]$. 
Note that $F_{l,m}=0$ if $l$ or $m$ are negative.
In the end, the polaron-transformed Hamiltonian is expressed in terms of the dressed states
\begin{equation}\label{Hall_dressedstate}
    \begin{split}
        \tilde{H}_\text{eff}(\{R\})
        &=\sum_{m}V_{\{D\}m}(\{R\})|\{D\}m\rangle\langle\{D\}m|\\&+\sum_{m}\sum_{j}V_{\{A_{j}\}m}(\{R\})|\{A_{j}\}m\rangle\langle\{A_{j}\}m|\\
         &+\sum_{l,m}\sum_{j}{\cal T}_{lm}|\{A_{j}\}l\rangle\langle\{D\}m|+\text{h.c.} 
    \end{split}
\end{equation}
where $\text{h.c.}$ denotes the Hermitian conjugate. 


Several physical implications can be obtained from this effective Hamiltonian.
We notice that the effective diabatic coupling in Eq.~\eqref{Hall_dressedstate} includes three channels for the electronic transition that occurs from $\{D\}$ to $\{A_{j}\}$: 
\begin{enumerate}
    \item For $l=m$, such as ${\cal T}_{00}|\{A_{j}\}0\rangle\langle\{D\}0|$, the effective diabatic coupling corresponds to electron transfer without involving photon absorption or emission.
    \item For $l<m$, such as ${\cal T}_{01}|\{A_{j}\}0\rangle\langle\{D\}1|$, the cavity photon is absorbed ($|1\rangle\rightarrow|0\rangle$) as the excess electron makes a transition ($\{D\}\rightarrow\{A_{j}\}$). 
    This ET process is usually considered a photon-mediated ET.
    \item For $l>m$, such as ${\cal T}_{10}|\{A_{j}\}1\rangle\langle\{D\}0|$, the excess electron makes a transition ($\{D\}\rightarrow\{A_{j}\}$) and emit a cavity photon ($|0\rangle\rightarrow|1\rangle$). 
    This ET process leads to ET-induced light emission.
\end{enumerate} 
Also note that, as a result of the polaron transformation, the overall effect of the dipole self-energy is included within the effective coupling ${\cal T}_{lm}$ in terms of $F_{l,m}$. 
For the limit case where $g_{AD}=0$, the matrix element of the shift operator becomes $F_{l,m}=\delta_{lm}$, so that ${\cal T}_{00}={\cal T}_{11}=H_{AD}$ recovers the standard ET through the diabatic coupling.
${\cal T}_{01}=\hbar\omega_c t_{AD}$, ${\cal T}_{10}=-\hbar\omega_c t_{AD}$ represents the ET coupled to the cavity photon through the transition dipole moment of the molecule.

In principle, the time evolution of the coupled electron-nuclei system governed by Eq.~\eqref{Hall_dressedstate} can be simulated using mixed quantum-classical dynamics. 
This simulation includes propagating $M\times(N+1)$ quantum states (where $M$ is the maximal number of photons) and integrating $N$-dimensional classical trajectories following the Langevin equation. 
While such simulations are possible for a few dozen molecules, the computational cost scales unfavorably with increasing $N$.

\section{Method}\label{sec:method}
With the effective Hamiltonian in hand, we now derive the electron transfer rate of a single donor-acceptor pair under the influence of the collective interactions through a shared photon mode. 
To this end, we implement the following techniques.
First, we utilize the timescale difference between the reorganization process and thermal fluctuation to focus on the electron transfer dynamics of a single donor-acceptor pair.
Second, we employ the rotating wave approximation (RWA) in the nearly-resonance conditions between the cavity photon frequency and the donor-acceptor energy gap $\hbar\omega_c\approx|E_{A}-E_{D}|$. 
Third, we account for the overall effect of the fluctuating local modes in terms of polariton relaxation.

\subsection{Reorganization--Fluctuation separation}
We observe that the effective Hamiltonian is a function of all the primary modes $\{R_j|j=1,\cdots,N\}$, which are coupled to the electronic transitions and the thermal bath of their corresponding molecules.
The main driving forces of the nuclear motion are:
(a) Reorganization: As the excess electron makes a transition from the donor to the acceptor  ($|D_j\rangle\rightarrow|A_j\rangle$), the equilibrium configuration of the corresponding nuclear mode will shift from the minimum of $V_D(R_j)$ to the minimum of $V_A(R_j)$, leading to reorganization of the corresponding primary mode.
(b) Thermal fluctuations: The Langevin dynamics of the primary mode leads to stochastic fluctuation of $R_j$. 
The time scale of thermal fluctuation is usually much faster than the reorganization process.
Due to this timescale difference, it is intuitively reasonable to imagine that, when a certain molecule goes through the reorganization process, all the other molecules are unaffected and follow thermal fluctuations.

With this observation, we now separate the reorganization process and thermal fluctuations.
Specifically, we focus on the reorganization process of a single molecule (say $j=1$) and treat the nuclear motions of the other $N-1$ molecules using the Langevin dynamics ($j=2,\cdots,N$). 
On the one hand, the local mode $R_1$ is considered the \textit{active} mode that is coupled to the electronic transition $|\{D\}\rangle\rightarrow|\{A_1\}\rangle$ and reorganize in response to the electronic transition.
On the other hand, the other local modes $\{R_j|j=2,\cdots,N\}$ are considered as a set of \textit{fluctuating} modes that are decoupled from the electronic transitions of their corresponding donor-acceptor pair.

To further simplify the system, we decompose the effective Hamiltonian into $\tilde{H}_\text{eff}(\{R\})=\tilde{H}_0(R_1)+\tilde{H}^\prime$.
Here the active-mode Hamiltonian is in the subspace of the active mode $R_1$
\begin{equation}\label{Hall_active}
    \begin{split}
        \tilde{H}_0(R_1)
        =&\sum_{m}V_{D,m}(R_1)|\{D\}m\rangle\langle\{D\}m|\\
        &+\sum_{m}V_{A,m}(R_1)|\{A_1\}m\rangle\langle\{A_1\}m|\\
        &+\sum_{m}\sum_{j=2}^{N}[V_{D,m}(R_1)+E_{AD}]|\{A_j\}m\rangle\langle\{A_j\}m|\\
         &+\sum_{l,m}\sum_{j}{\cal T}_{lm}|\{A_{j}\}l\rangle\langle\{D\}m|+\text{h.c.}\\
    \end{split}
\end{equation}
where we define $V_{D,m}(R_1)=V_{D}(R_1)+m\hbar\omega_c$ and $V_{A,m}(R_1)=V_{A}(R_1)+m\hbar\omega_c$.
Here we use $V_A(R_j)=E_{AD}+V_D(R_j)+\lambda_v{R_j}$ and $E_{AD}=E_A-E_D$ is the energy gap.
On the other hand, the fluctuation Hamiltonian is given in terms of the fluctuating modes $\{R_j|j=2,\cdots,N\}$
\begin{equation}\label{fluctuatingHamiltonian}
    \begin{split}
        \tilde{H}^\prime=&\sum_{j=2}^N V_D(R_j)\times \tilde{I}\\
        &+\sum_{m}\sum_{j=2}^N\lambda_v R_{j}|\{A_{j}\}m\rangle\langle\{A_{j}\}m|
    \end{split}
\end{equation}
where $\tilde{I}=\sum_{m}[|\{D\}m\rangle\langle\{D\}m|+\sum_{j}|\{A_j\}m\rangle\langle\{A_j\}m|]$ is the identity matrix.
Note that the fluctuating modes $\{R_j|j=2,\cdots,N\}$ follow the Langevin equation and the second term leads to stochastic modulation of the energy surface of the $|\{A_{j}\}m\rangle$ state. 

Based on this separation, we derive the Marcus ET rates for the active mode using $\tilde{H}_0(R_1)$ and estimate polariton relaxation using Fermi's golden rule (FGR) rate as induced by the fluctuating modes using $\tilde{H}^\prime$.

\subsection{Rotating wave approximations (RWA)}\label{subsec:RWA}
\begin{figure}
    \includegraphics[width=\linewidth]{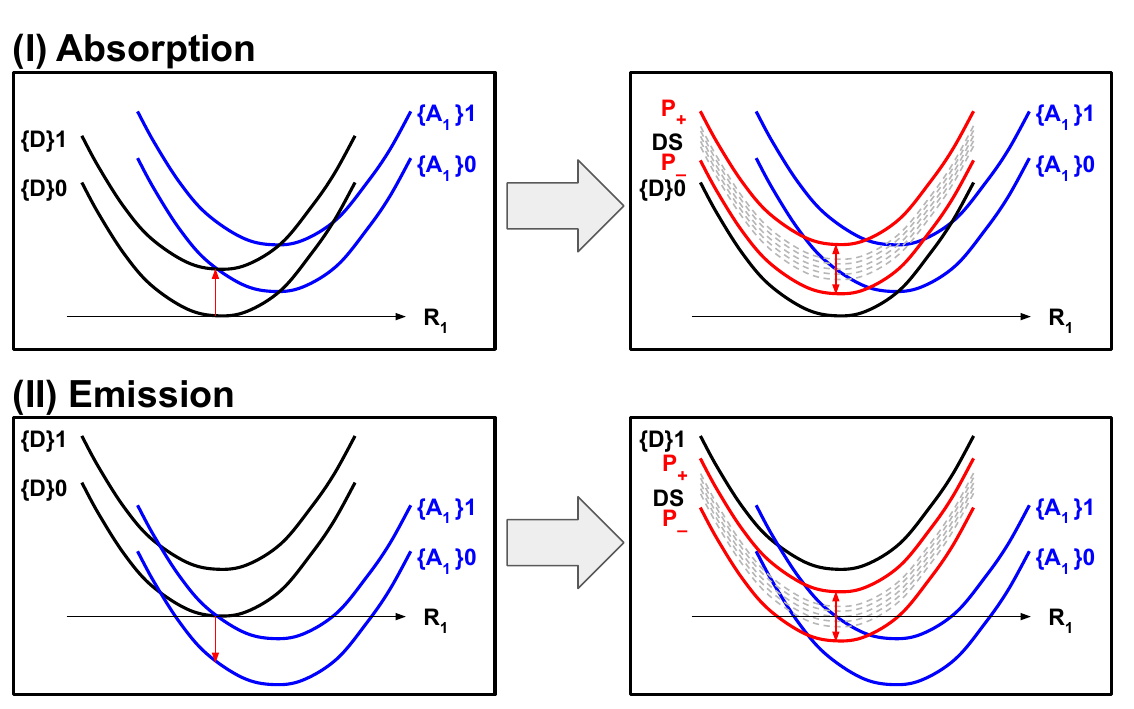}
    \caption{Schematic illustration of the potential energy surfaces in the active mode coordinate ($R_1$). The left panels are the $\{D\}m$ states (black) and the $\{A_1\}m$ states (blue) ($m=0,1$) for a single donor-acceptor pair. For (I), the PES of $\{D\}1$ and $\{A_1\}0$ is barrier-less, implying that the ET is coupled to photon absorption. For (II), the PES of $\{D\}0$ and $\{A_1\}1$ is barrier-less, implying that the ET is coupled to photon emission. 
    The right panels are the corresponding PES for $N$ donor-acceptor pairs. $P_{\pm}$ denotes the upper/lower polariton states (red) and $DS$ denotes the manifold of the dark states (grey dashed). Here $P_{\pm}$ are a combination of $\{D\}1$ and $\{A_2\}0\cdots\{A_N\}0$ for (I) and of $\{D\}0$ and $\{A_2\}1\cdots\{A_N\}1$ for (II). Note that the PES of $\{D\}m$ and $\{A_2\}m\cdots\{A_N\}m$ follows $V_D(R_1)$ in the $R_1$ coordinate. As a result, the Rabi splitting is proportional to $\sqrt{N-1}$. The dark states have thermal distribution due to Brownian motion.}
    \label{fig:PES_R1}
\end{figure}

In this section, we focus on the active-mode Hamiltonian $\tilde{H}^0(R_1)$ in the absence of the fluctuation Hamiltonian and assume that the cavity photon frequency and the donor-acceptor energy gap are nearly degenerate, i.e. $\hbar\omega_c\approx|E_{AD}|$.
Here we define the energy detuning to be $\Delta = \hbar\omega_c- |E_{AD}|$ and consider the following two conditions.
\begin{description}
    \item[(I) Absorption] When $E_{AD}>0$, $|\{D\}m+1\rangle$ and $|\{A_j\}m\rangle$ are nearly degenerate, i.e. the $D\rightarrow A$ electron transition is coupled to a cavity photon absorption, 
    \item[(II) Emission] When $E_{AD}<0$, $|\{D\}m\rangle$ and $|\{A_j\}m+1\rangle$ are nearly degenerate, i.e. the $D\rightarrow A$ electron transition is coupled to a cavity photon emission.
\end{description}
Fig.~\ref{fig:PES_R1} illustrates the potential energy surfaces in the $R_1$ coordinate. 
Note that the energy surfaces of the $\{D\}m$ and $\{A_{j}\}m$ (for $j\neq1$) follow $V_{D}(R_1)$ with a vertical energy shift, while the energy surfaces of $|\{A_{1}\}m\rangle$ follow $V_{A}(R_1)+m\hbar\omega_c$.
For simplicity, we restrict the subspace of photons to the vacuum ($m=0$) and single photon ($m=1$) states. 

\subsubsection{RWA-(I)}

For RWA-(I), we consider $E_{A}>E_{D}$ and the detuning is $\Delta = \hbar\omega_c- |E_{AD}|= \hbar\omega_c- E_{AD}$. 
In this case, $|\{D\}1\rangle$ and $|\{A_j\}0\rangle$ (for $j\neq1$) are nearly resonant and collectively coupled through the effective diabatic coupling term ${\cal T}_{01}$. 
Thus, the eigenstates within this subspace include the upper/lower polariton states  $|{P_\pm^{\text{(I)}}}\rangle$ and $N-2$ dark states (see Appendix~\ref{appendix_RWA}).
Since the dark states are spanned by $|\{A_j\}0\rangle$ (for $j\neq1$) (excluding $|\{D\}1\rangle$), we choose the initial state to be the donor state which does not populate the dark states.
That being said, the effect of the dark states as induced by the fluctuating modes will be included in terms of polariton relaxation in Sec.~\ref{subsec:relaxation}.

In the absence of the dark states, the active-mode Hamiltonian can be reduced and expressed in terms of the basis states $|\{D\}0\rangle$, $|\{A_1\}0\rangle$, $|{P_+^{\text{(I)}}}\rangle$, $|{P_-^{\text{(I)}}}\rangle$, $|\{A_1\}1\rangle$ in the matrix form
\begin{widetext}
\begin{equation}\label{H_RWAI_final}
\begin{split}
  \tilde{\mathbf{H}}_0^\text{(I)}(R_1) &=\frac{1}{2}\omega_v^2R_1^2+E_A\\ 
    &+\left[\begin{array}{ccccc}
    E_{DA} & {\cal T}_{00}^* & 0 & 0 & 0\\
    {\cal T}_{00} & \lambda_v R_{1} & {\cal T}_{01}\cos\theta^\text{(I)} & {\cal T}_{01}\sin\theta^\text{(I)} & 0\\
    0 & {\cal T}_{01}^{*}\cos\theta^\text{(I)} & \Omega^\text{(I)}_+ & {0} & {\cal T}_{11}^*\cos\theta^\text{(I)}\\
    0 & {\cal T}_{01}^{*}\sin\theta^\text{(I)} & {0} & \Omega^\text{(I)}_- & {\cal T}_{11}^*\sin\theta^\text{(I)}\\
    0 & 0 & {\cal T}_{11}^{*}\cos\frac{\theta_\text{I}}{2} & {\cal T}_{11}^{*}\sin\frac{\theta_\text{I}}{2} & \lambda_v R_{1}+\hbar\omega_{c}
    \end{array}\right]  
\end{split}
\end{equation}
\end{widetext}
Here we express $V_D(R_1)$ and $V_A(R_1)$ using Eq.~\eqref{V_D_and_V_A}.
The energy of the polariton state is given by 
\begin{equation}\label{polaritonenergy-I}
    \Omega^\text{(I)}_{\pm} = \frac{\Delta}{2}\pm\sqrt{\left(\frac{\Delta}{2}\right)^{2}+(N-1)|{\cal T}_{01}|^{2}}
\end{equation}
with respect to the eigenstate
\begin{equation}
\begin{split}
    |{P^\text{(I)}_{+}}\rangle &= \cos\theta^\text{(I)}|\{D\}1\rangle\\
    &+\sin\theta^\text{(I)}\sum_{j=2}^{N}\frac{1}{\sqrt{N-1}}|\{A_{j}\}0\rangle 
\end{split}
\end{equation}
\begin{equation}
\begin{split}
    |{P^\text{(I)}_{-}}\rangle &= \sin\theta^\text{(I)}|\{D\}1\rangle\\
    &-\cos\theta^\text{(I)}\sum_{j=2}^{N}\frac{1}{\sqrt{N-1}}|\{A_{j}\}0\rangle
\end{split}
\end{equation}
Here the Hopfield coefficients are defined in terms of the mixing angle, $\theta^\text{(I)}=\Theta({\cal T}_{01})$, where $\Theta$ is defined by 
\begin{align}\label{mixing_angle}
    \cos\Theta({\cal T}) &=\sqrt{\frac{1}{2}+\frac{1}{2}\frac{\Delta/2}{\sqrt{\left({\Delta}/{2}\right)^{2}+(N-1)|{\cal T}|^{2}}}} \\
    \sin\Theta({\cal T})  &=\sqrt{\frac{1}{2}-\frac{1}{2}\frac{\Delta/2}{\sqrt{\left({\Delta}/{2}\right)^{2}+(N-1)|{\cal T}|^{2}}}}
\end{align}
Note that the polariton states hybridize the cavity photon excitation $|\{D\}1\rangle$ and the electronic transitions $|\{A_j\}0\rangle$ ($j\neq1$), and the electronic transition of the active molecule (i.e $|\{A_{1}\}0\rangle$ state) is not involved in the polariton states. 
When the electronic transition is resonant with the cavity photon frequency (i.e. $\Delta=0$), the mixing angle becomes $\theta^\text{(I)}=\pi/4$ and the effective Rabi splitting is $ \Omega^\text{(I)}_{+}- \Omega^\text{(I)}_{-}=2\sqrt{(N-1)|{\cal T}_{01}|^{2}}$. 

\subsubsection{RWA-(II)}
For RWA-(II), we implement a similar analysis as in RWA-(I) for the case $E_{D}>E_{A}$. 
The detuning is $\Delta = \hbar\omega_c- |E_{AD}|=\hbar\omega_c-E_{DA}$. 
For this case, $|\{D\}0\rangle$ and $|\{A_j\}1\rangle$ (for $j\neq1$) are nearly resonant and collectively coupled via ${\cal T}_{10}$, the upper/lower polariton states $|{P_\pm^{\text{(II)}}}\rangle$. 
Excluding the dark state, the active-mode RWA-(II) Hamiltonian can be expressed in terms of the basis states  $|\{A_1\}0\rangle$, $|{P^\text{(II)}_{+}}\rangle$, $|{P^\text{(II)}_{-}}\rangle$ $|\{A_1\}1\rangle$, $|\{D\}1\rangle$ (see Appendix~\ref{appendix_RWA})
\begin{widetext}
\begin{equation}\label{H_RWAII_final}
\begin{split}
  \tilde{\mathbf{H}}^\text{(II)}_0(R_1) &=\frac{1}{2}\omega_v^2R_1^2+E_{A}+\hbar\omega_{c}\\ 
    &+\left[\begin{array}{ccccc}
    -\hbar\omega_{c}+\lambda_{v}R_{1}&{\cal T}_{00}^{*}\sin\theta^{\text{(II)}} &{\cal T}_{00}^{*}\cos\theta^{\text{(II)}} &0&0\\
    {\cal T}_{00}\sin\theta^{\text{(II)}} &\Omega^\text{(II)}_+&0&{\cal T}_{10}^{*}\sin\theta^{\text{(II)}}&0\\
    {\cal T}_{00}\cos\theta^{\text{(II)}}&0&\Omega^\text{(II)}_-&{\cal T}_{10}^{*}\cos\theta^{\text{(II)}}&0\\
    0&{\cal T}_{10}\sin\theta^{\text{(II)}}&{\cal T}_{10}\cos\theta^{\text{(II)}}&\lambda_{v}R_{1}&{\cal T}_{11}\\
    0&0&0&{\cal T}_{11}^{*}&E_{DA}
    \end{array}\right]  
\end{split}
\end{equation}
\end{widetext}
Here the energy of the polariton state is given by 
\begin{equation}\label{polaritonenergy-II} 
    \Omega^\text{(II)}_{\pm}(N) =-\frac{\Delta}{2}\pm\sqrt{\left(\frac{\Delta}{2}\right)^{2}+(N-1)|{\cal T}_{10}|^{2}}
\end{equation}
The eigenstates corresponding to the upper and lower polaritons are the polaritonic states, which take the form of
\begin{equation}
\begin{split}
    |{P^\text{(II)}_+}\rangle &= \sin\theta^{\text{(II)}}|\{D\}0\rangle\\
    &+\cos\theta^{\text{(II)}}\sum_{j=2}^{N}\frac{1}{\sqrt{N-1}}|\{A_{j}\}1\rangle
\end{split}
\end{equation}
\begin{equation}
\begin{split}
    |{P^\text{(II)}_-}\rangle &= \cos\theta^{\text{(II)}}|\{D\}0\rangle\\
    &-\sin\theta^{\text{(II)}}\sum_{j=2}^{N}\frac{1}{\sqrt{N-1}}|\{A_{j}\}1\rangle
\end{split}
\end{equation}
Here the mixing angle is $\theta^{\text{(II)}}=\Theta({\cal T}_{10})$ where $\Theta$ is given by Eq.~\eqref{mixing_angle}.

\subsection{Electron transfer (ET) rates}
To estimate the electron transfer rates of these paths, we employ the Marcus theory.\cite{marcus_relation_1989,nitzan_chemical_2006}
For convenience, we express the Marcus ET rate from quantum state $|i\rangle$ to $|f\rangle$ in terms of
\begin{equation}
    {\cal K}(V,E_a)=\sqrt{\frac{\pi}{E_{r}k_{B}T}}\frac{|V|^{2}}{2\hbar}\exp\left[-\frac{E_a^2}{4k_{B}TE_r}\right]
\end{equation}
where $V=\langle i|H|f\rangle $ is the coupling between the quantum states, $E_r$ is the reorganization energy, and $E_a=E_{fi}+E_r$ is the activation energy.
Here the effective reorganization energy is $E_r=\frac{\lambda_v^{2}}{2\omega_{v}^{2}}$ for the energy surfaces given in Eq.~\eqref{H_RWAI_final}.

For the RWA-(I) case, the ET rates from the polariton states are given by
\begin{subequations}\label{K_rate_I}
    \begin{align}
    {k}^\text{(I)}_{P_{+}\rightarrow\{A_1\}0}&={\cal K}({\cal T}_{01}\cos\theta^\text{(I)}, \Omega^\text{(I)}_{+})\\
    {k}^\text{(I)}_{P_{-}\rightarrow\{A_1\}0}&={\cal K}({\cal T}_{01}\sin\theta^\text{(I)}, \Omega^\text{(I)}_{-})\\
    {k}^\text{(I)}_{P_{+}\rightarrow\{A_1\}1}&={\cal K}({\cal T}_{11}\cos\theta^\text{(I)}, \Omega^\text{(I)}_{+}-\hbar\omega_c)\\
    {k}^\text{(I)}_{P_{-}\rightarrow\{A_1\}1}&={\cal K}({\cal T}_{11}\sin\theta^\text{(I)}, \Omega^\text{(I)}_{-}-\hbar\omega_c)
    \end{align}
\end{subequations}
The ET rate from $\{D\}0$ to $\{A_1\}0$ does not couple to the photon absorption and emission, ${k}^\text{(I)}_{\{D\}0\rightarrow\{A_1\}0}={\cal K}({\cal T}_{00},E_{AD})$, which recover the standard ET rate of a single donor-acceptor pair.
\footnote{Note that since the minimal value of $V_D(R_1)$ is $E_D$ and that of $V_A(R_1)$ is $E_A-E_r$, the activation energy is $E_a=E_{AD}-E_r$.
In the limit $g_{AD}=0$,  ${\cal T}_{00}$ is reduced to $H_{AD}$.}


For the RWA-(II) case, the ET rates from the polariton states are given by
\begin{subequations}\label{K_rate_II}
    \begin{align}
    {k}^\text{(II)}_{P_{+}\rightarrow\{A_1\}1}&={\cal K}({\cal T}_{10}\sin\theta^{\text{(II)}},\Omega^\text{(II)}_{+})\\
    {k}^\text{(II)}_{P_{-}\rightarrow\{A_1\}1}&={\cal K}({\cal T}_{10}\cos\theta^{\text{(II)}},\Omega^\text{(II)}_{-})\\
    {k}^\text{(II)}_{P_{+}\rightarrow\{A_1\}0}&={\cal K}({\cal T}_{00}\sin\theta^{\text{(II)}},\Omega^\text{(II)}_{+}+\hbar\omega_c)\\
    {k}^\text{(II)}_{P_{-}\rightarrow\{A_1\}0}&={\cal K}({\cal T}_{00}\cos\theta^{\text{(II)}},\Omega^\text{(II)}_{-}+\hbar\omega_c)
    \end{align}
\end{subequations}
The ET rate from $\{D\}1$ to $\{A_1\}1$ does not couple to the photon absorption and emission, ${k}^\text{(II)}_{\{D\}1\rightarrow\{A_1\}1}={\cal K}({\cal T}_{11},E_{DA})$. 
Note that, in the limit $g_{AD}=0$, the effective diabatic coupling is reduced to ${\cal T}_{00}={\cal T}_{11}=H_{AD}$.
Again, the ET rate recovers the typical electron transfer process for a single donor-acceptor pair.

\subsection{Polariton relaxation (PR) rate}\label{subsec:relaxation}
Now we turn our attention to the effect of the fluctuation Hamiltonian $\tilde{H}^\prime$ given by Eq.~\eqref{fluctuatingHamiltonian}. 
In the presence of the fluctuation Hamiltonian, the polariton states are coupled to the dark states, leading to polariton relaxation.
Here we treat the overall effect of the fluctuation Hamiltonian using the second-order time-dependent perturbation theory and the PR rate for $P_\pm$ can be estimated by the Fermi's golden rule
\begin{equation}
    \Gamma_\pm=\frac{1}{\hbar}\sum_{k=1}^{N-1} \int_{-\infty}^{\infty}dt \langle{P_\pm}| \tilde{H}_I^\prime(t)|{X_k}\rangle\langle{X_k}|\tilde{H}_I^\prime(0)|{P_\pm}\rangle
\end{equation}
Here $|{X_k}\rangle$ for $k=1,\cdots,N-2$ are the dark states (see Appendix~\ref{appendix_RWA}).

Next, to evaluate the time correlation of $\tilde{H}_I^\prime$, we derive the time correlation function of the fluctuating modes $R_k(t)R_k(0)$.
Since we assume that $R_k(t)$ follows Langevin dynamics, its spectral density follows a Brownian distribution
\begin{equation}\label{Brownian}
    {\cal J}(\omega)
    =\frac{2\gamma k_BT}{(\omega^2-\omega_v^2)^2+\gamma^2\omega^2}
\end{equation}
Thus, the analytical expression for the PR rate can be written as 
\begin{subequations}\label{PR-rate_I}
    \begin{align}
    \Gamma_+^\text{(I)}&=
    \frac{\lambda_v^2}{\hbar}{\cal J}(\Omega_+^\text{(I)})\sin^2\theta^\text{(I)}\\
    \Gamma_-^\text{(I)}&=
    \frac{\lambda_v^2}{\hbar}{\cal J}(\Omega_-^\text{(I)})\cos^2\theta^\text{(I)}
    \end{align}
\end{subequations}
for the absorption case, and 
\begin{subequations}\label{PR-rate_II}
    \begin{align}
    \Gamma_+^\text{(II)}&=
    \frac{\lambda_v^2}{\hbar}{\cal J}(\Omega_+^\text{(II)})\cos^2\theta^\text{(II)}\\
    \Gamma_-^\text{(II)}&=
    \frac{\lambda_v^2}{\hbar}{\cal J}(\Omega_-^\text{(II)})\sin^2\theta^\text{(II)}
    \end{align}
\end{subequations}
for the emission case.
This FGR rate can be comprehended as follows. 
In the fast modulation limit, the fluctuating modes $R_{k}(t)$ samples the entire Brownian distribution during the reorganization process of the active mode. 
Thus, the dark states can be approximated as a set of quasi-continuum states following the spectral distribution ${\cal J}(\omega)$ and coupled to the polariton states with coupling strength $\lambda_v/\sqrt{N}\cos\frac{\theta}{2}$ or $\lambda_v/\sqrt{N}\sin\frac{\theta}{2}$ respectively. 

\section{Result}\label{sec:result}
To show the dependence on the number of molecules, we employ the parameters of the ``slow electron, fast cavity'' case in Ref.~\citenum{semenov_electron_2019}. 
For the molecular system, we use $|E_{AD}|=1.0\ \text{eV}$ for the energy gap between the donor and the acceptor and $H_{AD}=245\ \text{cm}^{-1}\approx 0.03\ \text{eV}$ for the diabatic coupling.
We choose $E_r = 1.0\ \text{eV}$ for the reorganization energy, $\hbar\omega_v = 80.6\ \text{cm}^{-1}$ for the characteristic frequency of the vibrational primary mode. 
For the thermal bath, we choose $\gamma=0.065\ \text{s}^{-1}$ for the damping coefficient and $1/\beta=k_BT=0.025\ \text{eV}$ for the temperature.

We consider the cavity photon resonant with the donor-acceptor energy gap (i.e. $\hbar\omega_c = 1.0\ \text{eV}$). 
The radiative coupling between the electronic transition and the cavity photon is chosen to be $\hbar\omega_c |t_{AD}|=69\ \text{cm}^{-1}$ (i.e. $|t_{AD}|=8.5\times10^{-3}$).
The coupling parameter for the Frank-Condon shift operator is given by $|g_{AD}|=0.5$.  

For the resonance situation (i.e. $\Delta=0$), the effective Rabi splitting between the polariton states is given by Eq.~\eqref{polaritonenergy-I} and Eq.~\eqref{polaritonenergy-II}
\begin{align}
    \Omega^\text{(I)}_{\pm}(N) &= \pm\sqrt{(N-1)|{\cal T}_{01}|^{2}}\\
    \Omega^\text{(II)}_{\pm}(N) &= \pm\sqrt{(N-1)|{\cal T}_{10}|^{2}}
\end{align}
which increases with $N$. 
The mixing angle becomes $\cos\theta^{(\text{I})}=\sin\theta^{(\text{I})}=\frac{1}{\sqrt{2}}$ and $\cos\theta^{(\text{II})}=\sin\theta^{(\text{II})}=\frac{1}{\sqrt{2}}$, which does not depend on $N$.
Thus, the $N$-dependence of the ET rate is determined by the PES of the polariton states (see Fig.~\ref{fig:PES_R1}).

\begin{figure}
\includegraphics[width=0.48\linewidth]{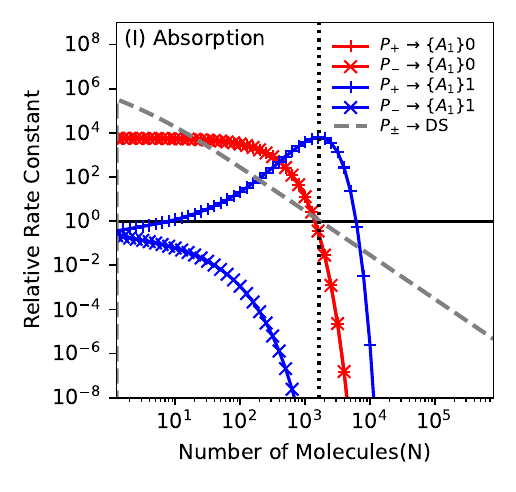}
\includegraphics[width=0.48\linewidth]{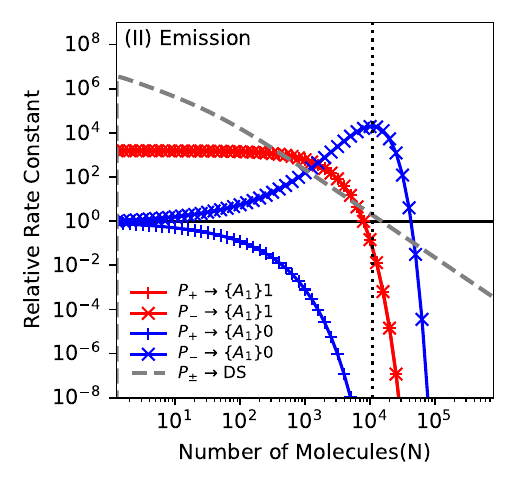}
        \caption{The relative rate constants for electron transfer (ET) process and polariton relaxation (PR) as a function of the number of the donor-acceptor pairs ($N$) for (I) absorption and (II) emission cases. 
        Here all the relative rate constant curves are 
        The ET rates of Path (b) (red solid lines) and Path (c) (blue solid lines) and the PR rates (grey dashed line) are plotted relative to the rate constant of Path (a) respectively ($\{D\}0\rightarrow\{A_1\}0$ for (I) and $\{D\}1\rightarrow\{A_1\}1$ for (II)). 
        We note that $P_{+}\rightarrow\{A_1\}1$ of (I) and $P_{-}\rightarrow\{A_0\}1$ of (II) exhibit non-monotonic $N$-dependence and predict a peak when the ET process is barrier-less.
        The black dotted line indicates the maximal relative rate $N^\text{(I)}_\text{max} = 1636$ and $N^\text{(II)}_\text{max} = 10785$. 
        As $N$ increases beyond $N_\text{max}$, both the ET and PR rates decrease and the PR rate dominates for large $N$. }\label{fig:ET/RP-N}
\end{figure}

\subsubsection{Analytical expressions for the ET and PR rates}
The ET paths are indicated as the horizontal arrows in Fig.~\ref{fig:PES_R1}.
Here we categorize the ET paths as follows:
\begin{itemize}
    \item Path (a): $\{D\}0\rightarrow\{A_1\}0$ for (I) and $\{D\}1\rightarrow\{A_1\}1$ for (II). The ET process is from the donor state (not polaritonic states) and does not change the cavity photon number. The ET rate constant does not depend on $N$ and recovers the standard ET rate in the absence of the cavity.
    \item Path (b): $P_\pm^{(\text{I})}\rightarrow\{A_{1}\}0$ and  $P_\pm^{(\text{II})}\rightarrow\{A_{1}\}1$. Note that $P_\pm^{(\text{I})}$ ($P_\pm^{(\text{II})}$) includes $\{D\}1$ ($\{D\}0$), so the ET process of Path (b) is coupled to photon absorption (I) and emission (II). 
    \item Path (c): $P_\pm^{(\text{I})}\rightarrow\{A_{1}\}1$ and $P_\pm^{(\text{II})}\rightarrow\{A_{1}\}0$. The ET process of Path (c) does not change the photon number. 
    Explicitly, the ET rates are given by Eq.~\eqref{K_rate_I} and Eq.~\eqref{K_rate_II}.
\end{itemize}
We notice that, for the resonance case, Path (c) predicts a turnover behavior as $N$ increases. Specifically, ${k}^\text{(I)}_{P_{+}\rightarrow\{A_1\}1}$ and ${k}^\text{(II)}_{P_{-}\rightarrow\{A_1\}0}$ has a maximal ET rate constant when $\Omega^\text{(I)}_+-\hbar\omega_c=0$ and $\Omega^\text{(II)}_0+\hbar\omega_c=0$ (i.e. the polariton PES becomes barrier-less) respectively. 
The maximal ET rate occurs at
\begin{subequations}\label{N_max_resonant}
\begin{align}
 N^\text{(I)}_\text{max} &= 1 + \frac{\hbar\omega_c}{|{\cal T}_{01}|^2} \\
 N^\text{(II)}_\text{max} &= 1 + \frac{\hbar\omega_c}{|{\cal T}_{10}|^2}
\end{align}
\end{subequations}
In the limit of large $N\rightarrow\infty$, both paths (b) and (c) decrease with $N$, and the asymptotic scaling of the ET rates is ${k}^\text{(I)}\propto{k}^\text{(II)}\propto e^{-N}$.

As far as the PR rates are concerned, since we assume the resonance condition and the Brownian distribution is symmetric, the relaxation rate is identical for the upper and lower polaritons, $\Gamma_+^\text{(I)}=\Gamma_-^\text{(I)}$ and $\Gamma_+^\text{(II)}=\Gamma_-^\text{(II)}$ 
Explicitly, the PR rates are given by Eq.~\eqref{PR-rate_I} and Eq.~\eqref{PR-rate_II}.
Here we emphasize that the PR rates decrease with $N$, and the asymptotic scaling is $\Gamma^\text{(I)}\propto\Gamma^\text{(II)}\propto N^{-2}$.

\subsubsection{$N$-dependence of the ET and PR rates}
In Fig.~\ref{fig:ET/RP-N}, we show the $N$ dependence of the ET and PR rate constants for the absorption and emission cases. 
To exhibit the modification as induced by the cavity photon, the ET rate constants of Path (b) and (c) and the PR rate are compared relative to Path (a) (i.e. divided by ${k}^\text{(I)}_{\{D\}0\rightarrow\{A_1\}0}$ and $k^\text{(II)}_{\{D\}1\rightarrow\{A_1\}1}$ respectively).

Regarding the ET rates, we observe the following features.
(\emph{i}) The ET rates of Path (b) are strongly enhanced compared to Path (a) because we choose $|E_{AD}|=\hbar\omega_c$ and the activation energy is zero when $N=1$.  
As $N$ increases, since the PESs of the $P_\pm$ shift away from the barrier-less condition (see Fig.~\ref{fig:PES_R1}), the ET rate of Path (b) drops exponentially. 
(\emph{ii}) For path (c), we observe non-monotonic $N$-dependence of the ET rate, specifically ${P_{+}^\text{(I)}\rightarrow\{A_1\}1}$ for the absorption case and ${P_{-}^\text{(II)}\rightarrow\{A_1\}0}$ for the emission case.
At $N^\text{(I)}_\text{max}$ and $N^\text{(II)}_\text{max}$, the activation energy of the corresponding path becomes zero (i.e. the ET path becomes barrier-less).
As the number of molecules increases beyond this point, the ET rate constant decreases exponentially with $N$.

As far as polariton relaxation is concerned, the PR rate follows the distribution given by Eq.~\eqref{Brownian} evaluated at the polariton energy ($\Omega_\pm(N)$). 
In the regime where the ET rate exceeds the PR rate, the ET from the polariton states is faster than the relaxation to the dark states, contributing to the enhancement of the overall ET as induced by the cavity photon.
For example, Fig.~\ref{fig:ET/RP-N}-(I) shows the contribution of $P_\pm\rightarrow\{A_1\}0$ between $20\lesssim{N}\lesssim1300$ and the contribution of $P_{+}\rightarrow\{A_1\}1$ between $190\lesssim{N}\lesssim7000$.
While the PR rate can be larger or smaller than the ET rate depending on $N$, the polariton relaxation always dominates the overall cavity-induced effect for large $N$. This is because the ET rates Path (b) and (c) decrease exponentially and become smaller than the PR rate ($\propto N^{-2}$) as $N\rightarrow\infty$.

\begin{figure}
    \includegraphics[width=0.48\linewidth]{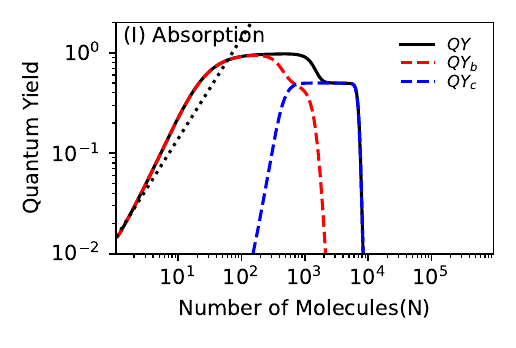}
    \includegraphics[width=0.48\linewidth]{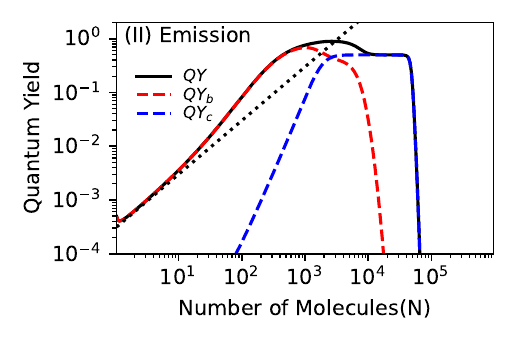}  
    \caption{Quantum yields as induced by photon-coupled ET are plotted as a function of $N$ for (I) absorption and (II) emission cases. The total quantum yield ($QY$, black solid line) can be divided into the contributions of Path (b) ($QY_b$, red dashed lines) and Path (c) ($QY_c$, blue dashed lines). For small $N$, we find that $QY_b\propto{N}$ (the dotted line), which agrees with the linear scaling of cavity-induced rate predicted by the Dicke model.\cite{semenov_electron_2019}. Note that $QY_b$ can reach $QY_b\approx1$ since both upper and lower polariton contribute to the ET process. $QY_c$ is limited by $QY_c<0.5$ as only one polaritonic state is involved in the ET.}
    \label{Yield vs N}
\end{figure} 

\subsubsection{Cavity-induced quantum yields ($QY$s)}
In order to quantify the competition mechanism between the ET and PR mechanisms, we use the equation of motion for the population and derive the quantum yield ($QY$) as induced by the cavity photon (see Appendix~\ \ref{appendix_yield}). 
Specifically, $QY$ can be calculated for Path (b) and (c)
\begin{subequations}
    \begin{align}
    {QY}_b^\text{(I)} &=
    \frac{{k}^\text{(I)}_{P_{+}\rightarrow\{A_1\}0}}{{K}^\text{(I)}_{+}} + 
    \frac{{k}^\text{(I)}_{P_{-}\rightarrow\{A_1\}0}}{{K}^\text{(I)}_{-}} \\ 
    {QY}_c^\text{(I)} &=
    \frac{{k}^\text{(I)}_{P_{+}\rightarrow\{A_1\}1}}{{K}^\text{(I)}_{+}} + 
    \frac{{k}^\text{(I)}_{P_{-}\rightarrow\{A_1\}1}}{{K}^\text{(I)}_{-}}
    \end{align}
\end{subequations}
where ${K}^\text{(I)}_{\pm}= {k}^\text{(I)}_{P_{\pm}\rightarrow\{A_1\}0}+{k}^\text{(I)}_{P_{\pm}\rightarrow\{A_1\}1}+\Gamma^\text{(I)}_{\pm}$. 
And for (II), we have 
\begin{subequations}
    \begin{align}
    {QY}_b^\text{(II)} &=
    \frac{{k}^\text{(II)}_{P_{+}\rightarrow\{A_1\}1}}{{K}^\text{(II)}_{+}} + 
    \frac{{k}^\text{(II)}_{P_{-}\rightarrow\{A_1\}1}}{{K}^\text{(II)}_{-}} \\
    {QY}_c^\text{(II)} &=
    \frac{{k}^\text{(II)}_{P_{+}\rightarrow\{A_1\}0}}{{K}^\text{(II)}_{+}} + 
    \frac{{k}^\text{(II)}_{P_{-}\rightarrow\{A_1\}0}}{{K}^\text{(II)}_{-}} 
    \end{align}
\end{subequations}
where ${K}^\text{(II)}_{\pm}= {k}^\text{(II)}_{P_{\pm}\rightarrow\{A_1\}0}+{k}^\text{(II)}_{P_{\pm}\rightarrow\{A_1\}1}+\Gamma^\text{(II)}_{\pm}$. 
Here the quantum yield is defined as the ratio of the final acceptor population (from the polaritonic states) to the cavity photon occupations corresponding to the path.
Note that the total quantum yield ($QY=QY_b+QY_c$) excludes the contribution of the standard ET (Path (a)) and accounts for only the acceptor population as included by the cavity effect.

In Fig.~\ref{Yield vs N}, several features of the cavity-induced quantum yield can be observed for both (I) and (II).
First, while Fig.~\ref{fig:ET/RP-N} shows that the ET rates of Path (b) are strongly enhanced for small $N$, $QY_b$ is relatively low because the PR rate is large.
More importantly, we find that the quantum yield scales linearly with $N$ in this regime, i.e. $QY_b\propto{N}$, which agrees with the linear dependence on $N$ predicted by the Dicke model.\cite{semenov_electron_2019}
Second, the quantum yield is significantly enhanced in the intermediate $N$ regime where the ET rates are larger than the PR rate. 
The enhancement can be attributed to the contributions of Path (b) and Path (c) separately. 
Third, we notice that the quantum yield of Path (b) can be as large as $QY_b\approx1$ since both $P_{\pm}$ contribute equally to the ET process. However, the quantum yield of Path (c) is limited to $QY_c\lesssim0.5$. This limitation is because the ET rates from $P^\text{(I)}_-$ and $P^\text{(II)}_+$ decrease monotonically with $N$ and are always much lower than the PR rates. Consequently, $QY_c$ is contributed only by one of the polaritonic states ($P^\text{(I)}_+$ or $P^\text{(II)}_-$). 
Finally, similar to Fig.~\ref{fig:ET/RP-N}, the quantum yields are suppressed for large $N$ since the PR rate dominates. 
As a result, we observe the suppression of the QY in the context of ETSC, which also manifests the large-$N$ issue.

\begin{figure}
    \includegraphics[width=0.49\linewidth]{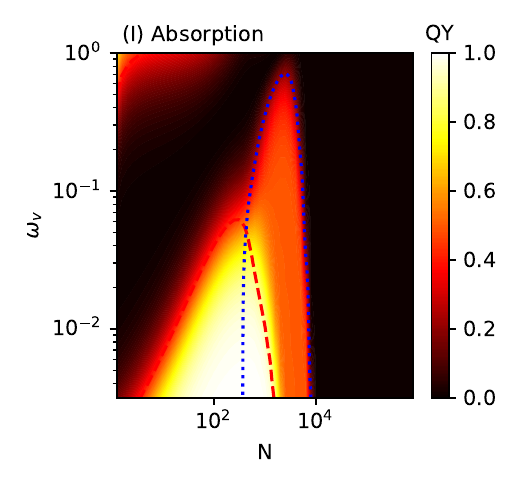}    
    \includegraphics[width=0.49\linewidth]{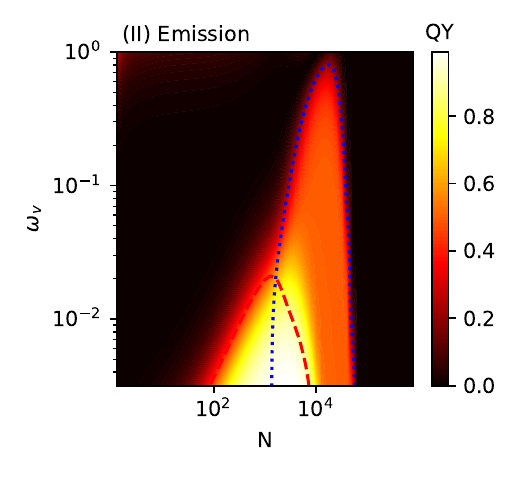}  
    \caption{The quantum yield with varying $\omega_v$ and $N$. The area enclosed by the red dashed line indicates the contribution of $QY_b$. The area enclosed by the blue dotted lines indicates the contribution of $QY_c$. We notice that $QY_b$ is suppressed in the fast bath regime and $QY_c$ is more robust against thermal fluctuation. }\label{fig:QY_Wvib}
\end{figure} 

\subsubsection{$\omega_v$ dependence of cavity-induced quantum yields}
Now we investigate the impact of the thermal bath on the cavity-induced quantum yields by varying the primary mode parameters.
In this subsection, we fix the reorganization energy $E_r=1\ \text{eV}$ and vary $\omega_v$ and $\lambda_v^2=2\omega_v^2$ (see Fig.~\ref{fig:QY_Wvib}).
In the slow bath regime (small $\omega_v$),  the QY enhancement regime of $N$ is enlarged, suggesting that the cavity-induced ET process becomes more effective. 
In the fast bath regime (large $\omega_v$), $QY_b$ is suppressed because the PR rate increase with $\omega_v$ ($\Gamma_\pm\propto \omega_v^2$ by Eq.~\eqref{Brownian}).

On the other hand, the contribution of Path (c) is more robust against the polariton relaxation in the fast bath regime.
Interestingly, we observe that the QY enhancement regime of $N$ for Path (c) does not change with $\omega_v$. 
To understand this, we recall that $N_\text{max}$ (Eq.~\ref{N_max_resonant}) is determined by the effective coupling ${\cal T}$ and the cavity photon frequency $\hbar\omega_c$, yet independent of $\omega_v$.
We emphasize that the polariton relaxation considered here is caused by the local molecular dynamics, rather than due to the cavity loss.

\section{Discussion and Conclusion}\label{sec:discussion}
Our investigation of the non-adiabatic electron transfer process in multiple donor-acceptor pairs within an optical cavity reveals several key insights into the interplay of collective excitation and local molecular dynamics. 
First, the upper and lower polaritons resulting from the collective interactions between electronic transitions of non-reacting molecules $\{A_j|j=2,\cdots,N\}$ and cavity photon excitation. This consideration accounts for the trend that the effect of the collective interaction on a single molecule should decrease with increasing $N$.
Second, the reacting molecule (molecule $1$ in this paper) is excluded from the formation of polariton and its electron-nuclei dynamics follows an effective spin-boson model with effective diabatic coupling and polariton-modulated energy surfaces.
Third, the nuclear motion of the non-reacting molecules is considered thermal fluctuations and causes polariton relaxation. Note that this relaxation is induced by intrinsic nuclear dynamics, rather than due to the cavity photon loss.
Finally, the cavity-induced quantum yield results from the interplay between electron transfer and polariton relaxation, showing non-monotonic dependence on $N$. We find linear scaling of the quantum yield for small $N$ (which agrees with extrapolation from a single molecule model) and exponential suppression for large $N$ (as the polariton relaxation dominates). 
This behavior is consistent with observations in polariton chemistry, where too many molecules in the cavity lead to a diminishing enhancement effect.\cite{chen_exploring_2024}

In conclusion, we provide an analytical framework for understanding the non-adiabatic electron transfer process in the context of strong light-matter coupling within an optical cavity. 
The easy-to-use formula for the electron transfer rate constant and polariton relaxation rate could guide the optimal conditions for cavity-enhanced chemical reactivity. 
Future work could build upon this framework to explore a broader range of molecular systems and cavity properties, for example the influence of cavity loss, inhomogeneous disorder of the molecular ensemble, and multi-photon excitations. Furthermore, more detailed dynamical properties can be obtained by employing non-adiabatic dynamics simulation methods, such as Floquet surface hopping and quasi-diabatic propagation,\cite{fiedlschuster_surface_2017,zhou_nonadiabatic_2020,hu_quasi-diabatic_2022} paving the way for more efficient design of polaritonic devices and materials.

\appendix


\section{Two flavors of rotating wave approximations}\label{appendix_RWA}


The matrix form of $\tilde{H}_0(R_1)$ can be written as
\begin{equation}
    \tilde{\mathbf{H}}_0(R_1)=\left[\begin{array}{ccc}
\mathbf{V}(R_1)+\mathbf{T}_{00} & \mathbf{T}_{01} & \cdots \\
\mathbf{T}_{01}^\dagger & \mathbf{V}(R_1)+\mathbf{T}_{11}+\hbar\omega_c\mathbf{I}  & \cdots \\
\vdots & \vdots  & \ddots \\
\end{array}\right]
\end{equation}
where $\mathbf{I}$ is the identity matrix of size $N+1$ and 
\begin{equation}
    \mathbf{V}_m(R_1)=\left[\begin{array}{ccccc}
V_{D}(R_1) &&&& \\
& V_{A}(R_1) & & & \\
& & \ddots &  & \\
& & & V_{D}(R_1)+E_{AD} & \\
& & & & \ddots 
\end{array}\right]
\end{equation}
\begin{equation}
    \mathbf{T}_{lm}=\left[\begin{array}{cccc}
0 & \cdots & {\cal T}_{ml}^* & \cdots \\
\vdots & & & \\
{\cal T}_{lm} & & & \\
\vdots & & & 
\end{array}\right] 
\end{equation}
Note that the basis states for the matrix form are in the order of $|\{D\}0\rangle$,  $|\{A_j\}0\rangle$ ($j=1,\cdots,N$), $|\{D\}1\rangle$,  $|\{A_j\}1\rangle$ ($j=1,\cdots,N$).
\paragraph{RWA-I}
For the case that $E_{AD}\approx\hbar\omega_c$,  we can employ the following approximations:
(i) Since $|\{D\}1\rangle$ is nearly resonant with $|\{A_j\}0\rangle$, we keep the ${\cal T}_{01}$ coupling terms and neglect the ${\cal T}_{10}$ coupling terms. 
(ii) Since the energy surfaces $|\{D\}0\rangle$ and $|\{A_j\}0\rangle$ (for $j\neq1$) are separated and also do not cross in the $R_1$ coordinate, we exclude the  ${\cal T}_{00}$ coupling terms, except for  ${\cal T}_{00}|\{D\}0\rangle\langle\{A_1\}0|+h.c.$. 
(iii) Similar to (ii),  we can also exclude the  ${\cal T}_{11}$ coupling terms, except for  ${\cal T}_{11}|\{D\}1\rangle\langle\{A_1\}1|+h.c.$ term, then eliminate the diagonal elements of $|\{A_j\}1\rangle$ (for $j\neq1$)  as they are not coupled to other states.
By doing so, we can write the RWA Hamiltonian of the first condition (RWA-I) in the matrix form:
\begin{widetext}
\begin{equation}
\begin{split}
    \tilde{\mathbf{H}}^\text{(I)}_0(R_1) = 
    \left[\begin{array}{ccccccc}
    V_{D}(R_{1}) & {\cal T}_{00}^* &  & & & &\\
    {\cal T}_{00} & V_{A}(R_{1}) &  &  &  & {\cal T}_{01} & \\
    &  & \ddots &  &  & \vdots & \\
    &  &  & V_{D}(R_{1})+E_{AD} &  & {\cal T}_{01} &\\
    &  &  &  & \ddots & \vdots & \\
    & {\cal T}_{01}^{*} & \cdots & {\cal T}_{01}^{*} & \cdots & V_{D}(R_{1})+\hbar\omega_{c} & {\cal T}_{11}^* \\
     &  &  &  &  & {\cal T}_{11} & V_{A}(R_{1})+\hbar\omega_{c}
    \end{array}\right]
\end{split}
\end{equation}
\end{widetext}
Here the basis states for the matrix form are in the order of $|\{D\}0\rangle$, $|\{A_1\}0\rangle$, $|\{A_j\}0\rangle$ ($j=2,\cdots,N$), $|\{D\}1\rangle$, $|\{A_1\}1\rangle$. 
Next, we diagonalize the submatrix spanned by the nearly-degenerate states $|\{D\}1\rangle$ and $|\{A_j\}0\rangle$ for $j=2,\cdots,N$, 
\begin{equation}\label{RWA-I_submatrix}
\begin{split}
    V_{D}(R_{1})+E_{AD}+\left[\begin{array}{cccc}
    \ddots &  &  & \vdots \\
     & 0 &  & {\cal T}_{01}\\
     &  & \ddots & \vdots \\
     \cdots & {\cal T}_{01}^{*} & \cdots & \Delta
    \end{array}\right]
\end{split}
\end{equation}
where $\Delta= \hbar\omega_c-|E_{AD}|=\hbar\omega_c-E_{AD}
$ is the energy detuning.
Finally, we obtain the polariton eigenvalues $\Omega^{\text{(I)}}_{\pm}(N)$ in Eq.~\eqref{polaritonenergy-I} and their corresponding eigenstates  $|{P_\pm^{\text{(I)}}}\rangle$.

In addition to the polaritonic states, diagonalizing the submatrix also results in $N-2$  degenerate eigenvalues. 
These degenerate states are considered \textit{dark}  due to their composition as a linear combination of dressed states sharing identical photon numbers
\begin{equation}
    |X^\text{(I)}_{k}\rangle=\sum_{j=2}^{N}x_{jk}|\{A_{j}\},0\rangle \\
\end{equation}
for $k=1,\cdots,N-2$. 
Here the coefficients follow $\sum_{j=2}^{N}x_{jk}=0$  so that the dark states do not couple to $|\{D\}1\rangle$.
We choose the dark state coefficient following Ref.~\citenum{heller_semiclassical_2018},
\begin{equation}
    x_{jk}=\frac{1}{\sqrt{N(N-1)}}-\sqrt{\frac{N-1}{N}}\delta_{jk}
\end{equation}
Since we consider the initial state of the electronic system to be on the donor state $\{D\}$, the dark states are not populated initially and the polariton states are not coupled to the dark states in the absence of the fluctuating Hamiltonian.
Therefore, the unperturbed RWA-I Hamiltonian can be expanded in terms of  $|\{D\}0\rangle$, $|\{A_1\}0\rangle$, $|{P_+^{\text{(I)}}}\rangle$, $|{P_-^{\text{(I)}}}\rangle$, $|\{A_1\}1\rangle$. 

\paragraph{RWA-II}
For the case that $E_{D}>E_{A}$ and $E_{D}-E_{A}=E_{DA}\approx\hbar\omega_c$, we follow a similar procedure and make the following approximations:
(i) Since $|\{D\}0\rangle$ is nearly resonant with $|\{A_j\}1\rangle$, we keep the ${\cal T}_{10}$ coupling terms and neglect the ${\cal T}_{01}$ coupling terms. 
(ii) Since the energy surfaces of  $|\{D\}1\rangle$ and $|\{A_j\}1\rangle$ (for $j\neq1$) are separated and also do not cross in the $R_1$ coordinate, we exclude the  ${\cal T}_{11}$ coupling terms, except for  ${\cal T}_{11}|\{D\}1\rangle\langle\{A_1\}1|$. 
(iii) Similar to (ii),  we can also exclude the  ${\cal T}_{00}$ coupling terms, except for ${\cal T}_{00}|\{D\}0\rangle\langle\{A_1\}0|+h.c.$ such that $|\{A_j\}0\rangle$ (for $j\neq1$)  states are not coupled to other states.
With these approximations, we can reduce the effective Hamiltonian as:
\begin{widetext}
\begin{equation}
\begin{split}
    \tilde{\mathbf{H}}^\text{(II)}_0(R_1) = 
    \left[\begin{array}{ccccccc}
    V_{A}(R_{1}) & {\cal T}_{00} &  & & & &\\
    {\cal T}_{00}^* & V_{D}(R_{1}) & \cdots & {\cal T}_{10}^* & \cdots & {\cal T}_{10}^* & \\
    & \vdots & \ddots &  &  &  & \\
    & {\cal T}_{10} &  & V_{D}(R_{1})+E_{AD}+\hbar\omega_c &  &  &\\
    & \vdots &  &  & \ddots & & \\
    & {\cal T}_{10} &  &  &  & V_{A}(R_{1})+\hbar\omega_{c} & {\cal T}_{11} \\
     &  &  &  &  & {\cal T}_{11}^* & V_{D}(R_{1})+\hbar\omega_{c}
    \end{array}\right]
\end{split}
\end{equation}
\end{widetext}
Here the basis states of the matrix format are rearranged in the order of $|\{A_1\}0\rangle$, $|\{D\}0\rangle$, $|\{A_j\}0\rangle$ ($j=2,\cdots,N$), $|\{A_1\}1\rangle$, $|\{D\}1\rangle$.
Next, we can diagonalize the submatrix 
\begin{equation}\label{RWA-II-submatrix}
\begin{split}
    V_{D}(R_{1}) + E_{AD}+\hbar\omega_c+
    \left[\begin{array}{cccc}
    -\Delta & \cdots & {\cal T}_{10}^* & \cdots \\
    \vdots &  &  & \\
    {\cal T}_{10} &  & &\\
    \vdots &  &  & 
    \end{array}\right]
\end{split}
\end{equation}
where the energy detuning is  $\Delta=\hbar\omega_c-|E_{AD}|=\hbar\omega_c-E_{DA}$. 
We then obtain the two eigenvalues $\Omega^{\text{(II)}}_{\pm}(N)$ in Eq.~\eqref{polaritonenergy-II}  and the corresponding eigenstates  $|{P_\pm^{\text{(II)}}}\rangle$.
The dark states in RWA(II) can be expressed as a linear combination of $|\{A_{j}\},1\rangle $
\begin{equation}
    |X^\text{(II)}_{k}\rangle=\sum_{j=2}^{N}x_{jk}|\{A_{j}\},1\rangle \\
\end{equation}
for $k=1,\cdots,N-2$.
Here we choose the same dark state coefficients as in RWA(I).

\section{Derivation of the golden rule relaxation rate}

To account for the relaxation as induced by the fluctuation Hamiltonian $\tilde{H}^\prime$, we employ Fermi's golden rule  (FGR).
We consider the polariton states $|P_{\pm}\rangle$ are coupled to a set of dark states from an initial state $|X_k\rangle$ for $k=1,\cdots,N-2$, then the relaxation rate can be estimated by 
\begin{equation}
    \Gamma_\pm=\frac{1}{\hbar}\sum_k \int_{-\infty}^{\infty}dt \langle{P_\pm}| \tilde{H}_I^\prime(t)|{X_k}\rangle\langle{X_k}|\tilde{H}_I^\prime(0)|{P_\pm}\rangle
\end{equation}
where $\tilde{H}_I^\prime(t)= e^{i\tilde{H}_0t/\hbar}\tilde{H}^\prime e^{-i\tilde{H}_0t/\hbar}$ is the interaction representation of the fluctuation Hamiltonian. 
Using the dark state expressions in Appendix \ref{appendix_RWA}, the matrix elements are 
\begin{equation}
\begin{split}
    \langle P^\text{(I)}_{+}|\tilde{H}_I^\prime(t)|X_{k}^\text{(I)}\rangle
    &=e^{i\Omega_+^\text{(I)}t/\hbar}\sin\theta^\text{(I)}\sum_{j=2}^{N}\frac{\lambda_v R_{j}(t)x_{jk}}{\sqrt{N-1}}\\
    &=-\frac{\lambda_v}{\sqrt{N}}e^{i\Omega_+^\text{(I)}t/\hbar}R_{k}(t)\sin\theta^\text{(I)}
\end{split}
\end{equation}
\begin{equation}
\begin{split}
    \langle P^\text{(I)}_{-}|\tilde{H}_I^\prime(t)|X_{k}^\text{(I)}\rangle
    &=-\cos\theta^\text{(I)}e^{i\Omega_-^\text{(I)}t/\hbar}\sum_{j=2}^{N}\frac{\lambda_v R_{j}(t)x_{jk}}{\sqrt{N-1}}\\
    &=\frac{\lambda_v}{\sqrt{N}}e^{i\Omega_-^\text{(I)}t/\hbar}R_{k}(t)\cos\theta^\text{(I)}
\end{split}
\end{equation}
\begin{equation}
\begin{split}
    \langle P^\text{(II)}_{+}|\tilde{H}_I^\prime(t)|X_{k}^\text{(II)}\rangle
    &=\cos\theta^\text{(II)}e^{i\Omega_+^\text{(II)}t/\hbar}\sum_{j=2}^{N}\frac{\lambda_v R_{j}(t)x_{jk}}{\sqrt{N-1}}\\
    &=-\frac{\lambda_v}{\sqrt{N}}e^{i\Omega_+^\text{(II)}t/\hbar}R_{k}(t)\cos\theta^\text{(II)}
\end{split}
\end{equation}
\begin{equation}
\begin{split}
    \langle P^\text{(II)}_{-}|\tilde{H}_I^\prime(t)|X_{k}^\text{(II)}\rangle
    &=-\sin\theta^\text{(II)}e^{i\Omega_-^\text{(II)}t/\hbar}\sum_{j=2}^{N}\frac{\lambda_v R_{j}(t)x_{jk}}{\sqrt{N-1}}\\
    &=\frac{\lambda_v}{\sqrt{N}}e^{i\Omega_-^\text{(II)}t/\hbar}R_{k}(t)\sin\theta^\text{(II)}
\end{split}
\end{equation}
Note that we use  $\sum_{j=2}^N R_j(t)=0$. 
We notice that the key element in the FGR rate expression is the Fourier transformation of the time correlation function 
\begin{equation}
    {\cal J}(\omega)= \int_{-\infty}^{\infty}dt e^{i\omega t/\hbar}\langle{R_k(t)}{R_k(0)}\rangle
\end{equation}
Suppose the timescale of the fluctuating mode is much faster than the reorganization process of the active mode, we can resolve the fluctuating mode ${R}_j(t)$ into its spectral components $r_{j,\ell}=\frac{1}{\tau_{R}}\int_{0}^{\tau_{R}}R_{j}(t)e^{-i\omega_{\ell}t}dt$. 
Here $\{r_{j,\ell}\}$ is the strength of the spectral component associated with frequency $\omega_{\ell}$.
Since the stochastic random force $\eta (t)$ is assumed to be white noise, the spectral distribution of $\{r_{j,\ell}\}$ follows the Brownian spectral density\cite{nitzan_chemical_2006} 
\begin{equation}\label{Brownian_distribution}
    \frac{1}{2\pi}\int_{-\infty}^{\infty}dt e^{-i\omega_\ell t} \langle{R_k(t)}{R_k(0)}\rangle
    =\frac{\gamma k_BT}{\pi}\frac{1}{(\omega_\ell^2-\omega_v^2)^2+\gamma^2\omega_\ell^2}
\end{equation}
So the Fourier transform of the time correlation function is
\begin{equation}
    {\cal J}(\Omega)= \frac{2\gamma k_BT}{(\Omega^2-\omega_v^2)^2+\gamma^2\Omega^2}
\end{equation}
Note that the spectral distribution does not depend on $k$, i.e. $\sum_k\rightarrow N-1$.

For the absorption case, the FGR relaxation rates are 
\begin{equation}
    \Gamma_+^\text{(I)}=(N-1)\frac{\lambda_v^2}{\hbar N}{\cal J}(\Omega_+^\text{(I)})\sin^2\theta^\text{(I)} \rightarrow 
    \frac{\lambda_v^2}{\hbar}{\cal J}(\Omega_+^\text{(I)})\sin^2\theta^\text{(I)}
\end{equation}
\begin{equation}
    \Gamma_-^\text{(I)}=(N-1)\frac{\lambda_v^2}{\hbar N}{\cal J}(\Omega_-^\text{(I)})\cos^2\theta^\text{(I)} \rightarrow 
    \frac{\lambda_v^2}{\hbar}{\cal J}(\Omega_-^\text{(I)})\cos^2\theta^\text{(I)}
\end{equation}
For the emission case, the FGR relaxation rates are  
\begin{equation}
    \Gamma_+^\text{(II)}=(N-1)\frac{\lambda_v^2}{\hbar N}{\cal J}(\Omega_+^\text{(II)})\cos^2\theta^\text{(II)} \rightarrow 
    \frac{\lambda_v^2}{\hbar}{\cal J}(\Omega_+^\text{(II)})\cos^2\theta^\text{(II)}
\end{equation}
\begin{equation}
    \Gamma_-^\text{(II)}=(N-1)\frac{\lambda_v^2}{\hbar N}{\cal J}(\Omega_-^\text{(II)})\sin^2\theta^\text{(II)} \rightarrow 
    \frac{\lambda_v^2}{\hbar}{\cal J}(\Omega_-^\text{(II)})\sin^2\theta^\text{(II)}
\end{equation}

\section{Derive quantum yields  from the rate equation}\label{appendix_yield}
With the ET and PR rates given in the main text, we can write the equation of motion for the population $\rho(t)$
\begin{align}
    \dot{\rho}_{0} &= -k_{0\rightarrow1}\rho_0\\
    \dot{\rho}_{1} &= k_{2\rightarrow1}\rho_2 + k_{3\rightarrow1}\rho_3 + k_{0\rightarrow1}\rho_0\\
    \dot{\rho}_{2} &= -k_{2\rightarrow1}\rho_2 -k_{2\rightarrow4}\rho_2 - \Gamma_+ \rho_2 \\
    \dot{\rho}_{3} &= -k_{3\rightarrow1}\rho_3 -k_{3\rightarrow4}\rho_3 - \Gamma_- \rho_3 \\
    \dot{\rho}_{4} &= k_{2\rightarrow4}\rho_2 + k_{3\rightarrow4}\rho_3 + k_{5\rightarrow4}\rho_5\\
    \dot{\rho}_{5} &= -k_{5\rightarrow4}\rho_5
\end{align}
For convenience, we denote the indices $0=\{D\}0$, $1=\{A_1\}0$, $2=P_+$, $3=P_-$, $4=\{A_1\}1$, $5=\{D\}1$,.
Note that $\rho_5=0$ and $k_{5\rightarrow4}=0$ for Case (I), and $\rho_0=0$ and $k_{0\rightarrow1}=0$ for Case (II).

The final populations are $\rho_{0}(\infty)=\rho_{2}(\infty)=\rho_{3}(\infty)=\rho_{5}(\infty)=0$ and 
\begin{align}
    \rho_{1}(\infty) &=\rho_0(0) + \frac{k_{2\rightarrow1}\rho_2(0)}{k_+}+ \frac{k_{3\rightarrow1}\rho_3(0)}{k_-} \\
    \rho_{4}(\infty) &=\rho_5(0) + \frac{k_{2\rightarrow4}\rho_2(0)}{k_+}+ \frac{k_{3\rightarrow4}\rho_3(0)}{k_-}
\end{align}
where $k_+= k_{2\rightarrow1} + k_{2\rightarrow4} + \Gamma_+$ and $k_-= k_{3\rightarrow1} + k_{3\rightarrow4} + \Gamma_-$. 
Here the first terms ($\rho_0(0)$ and $\rho_5(0)$) are contributed by the standard ET process (i.e Path (a)). 
In the following, we exclude the contribution of Path (a) to focus on the cavity-induced ET.

For the resonance case considered here, the initial UP/LP populations are the cavity photon occupation, i.e. $\rho_2(0)=\rho_3(0)=\rho_\text{cav}(m)$ where $m=1$ for (I) and $m=0$ for (II). 
Thus, we can define the quantum yield ($QY$) for Path (b) and (c) as follows. 
For (I), $QY_b$ corresponds to the final yield in the state $1=\{A_1\}0$ and $QY_c$ corresponds to the final yield in the state $4=\{A_1\}1$:
\begin{align}
    QY^\text{(I)}_b=\frac{\rho_{1}(\infty)}{\rho_\text{cav}(1)} &=\frac{k_{2\rightarrow1}}{k_+}+ \frac{k_{3\rightarrow1}}{k_-} \\
    QY^\text{(I)}_c=\frac{\rho_{4}(\infty)}{\rho_\text{cav}(1)} &=\frac{k_{2\rightarrow4}}{k_+}+ \frac{k_{3\rightarrow4}}{k_-}
\end{align}
For (II), $QY_b$ corresponds to the final yield in the state $4=\{A_1\}1$ and $QY_c$ corresponds to the final yield in the state $1=\{A_1\}0$:
\begin{align}
    QY^\text{(II)}_b=\frac{\rho_{4}(\infty)}{\rho_\text{cav}(0)} &=\frac{k_{2\rightarrow4}}{k_+}+ \frac{k_{3\rightarrow4}}{k_-}\\
    QY^\text{(II)}_c=\frac{\rho_{1}(\infty)}{\rho_\text{cav}(0)} &=\frac{k_{2\rightarrow1}}{k_+}+ \frac{k_{3\rightarrow1}}{k_-}
\end{align}

\bibliography{reference}

\end{document}